\documentclass[12pt,aps,ams,amsfonts,nofootinbib]{revtex4}

\usepackage{lineno,hyperref}
\modulolinenumbers[10]

\usepackage[T1]{fontenc}
\usepackage{textcomp}
\usepackage{bbm}
\usepackage{graphicx}
\usepackage{amssymb}
\usepackage{subfigure}
\usepackage{stackrel}
\usepackage{enumitem}
\usepackage{amsmath}
\usepackage{amsfonts}
\usepackage{bm}
\usepackage{color}
\usepackage{verbatim} %for comments
\usepackage{tcolorbox}
\usepackage[normalem]{ulem}
\usepackage{hyperref}
\usepackage{soul}

\newcommand*\diff{\mathop{}\!\mathrm{d}}

\newlist{subquestion}{enumerate}{1}
\setlist[subquestion,1]{label=(\alph*)}

\newcommand{\vect}[1]{\ensuremath{\bm{{#1}}}}
\newcommand{\ket}[1]{\ensuremath{\left|{#1}\right\rangle}}
\newcommand{\bra}[1]{\ensuremath{\left\langle{#1}\right |}}

\newcommand{\Trr}[1]{\textrm{Tr}\left[#1\right]}
\newcommand{\TrP}[2]{\textrm{Tr}_{#1}\left[#2\right]}

\newcommand{\beq}{\begin{equation}}
\newcommand{\eeq}{\end{equation}}
\newcommand{\bse}{\begin{subequations}}
	\newcommand{\ese}{\end{subequations}}\newcommand{\bea}{\begin{eqnarray}}
\newcommand{\eea}{\end{eqnarray}}
\newcommand{\bit}{\begin{itemize}}
	\newcommand{\eit}{\end{itemize}}
\newcommand{\bpmatrix}{\begin{pmatrix}}
	\newcommand{\epmatrix}{\end{pmatrix}}

\newcommand{\be}{\begin{equation}}
\newcommand{\ee}{\end{equation}}
\newcommand{\ben}{\begin{eqnarray}}
\newcommand{\een}{\end{eqnarray}}

\begin{document}

\title{Revisiting maximal average fidelity of teleportation}

\author{D. G. Bussandri$^{1}$, M. Portesi$^{1,2}$, A. P. Majtey$^{3,4}$}

\affiliation{$^1$Instituto  de  F\'isica  La  Plata  (IFLP),  CONICET, Diag. 113 e/ 63 y 64, 1900 La Plata, Argentina}
\affiliation{$^2$Facultad  de Ciencias Exactas, Universidad Nacional de La Plata, C.C. 67, 1900 La Plata, Argentina}
\affiliation{$^3$Instituto de F\'isica Enrique Gaviola (IFEG), CONICET}
\affiliation{$^4$Facultad de Matem\'atica, Astronom\'{\i}a, F\'{\i}sica y Computaci\'on, Universidad Nacional de C\'ordoba, Av. Medina Allende s/n, Ciudad Universitaria, X5000HUA C\'ordoba, Argentina}

\begin{abstract}
We obtain the maximal average fidelity corresponding to the standard quantum teleportation protocol for an arbitrary isotropic distribution of input states and an arbitrary resource state. We extend this result to a family of von Neumann measurements, which includes the projections onto the computational and Bell basis, considering a Bell-diagonal resource state.

We focus on three specific isotropic distributions of input states: 1) completely mixed input states, 2) states with a certain (fixed) degree of purity, and 3)  quasi-pure input states. We show that the standard quantum teleportation protocol can teleport arbitrary mixed states with higher average fidelity than its classical counterpart even when the resource of the protocol is a non-entangled state, specifically, a separable Werner state. Moreover, we find that the maximum average fidelity obtained with classical-quantum states used as a resource in a standard teleportation protocol also exceeds the classical fidelity. To establish the role played by the presence or absence of quantum correlations in the resource state and their relation with the correlations present in the von Neumann measurement performed by Alice, we analyze in detail the case of Bell diagonal resource states employing a family of monoparametric basis for which both the Bell and the computational (non-correlated) basis are included. Only in the case where the basis on which Alice measures is completely uncorrelated (computational basis) the maximum average fidelity does not exceed the classical fidelity for any resource state. In all other cases, the maximum average fidelity exceeds the classical one for a certain range of parameters describing the resource state, evidencing the importance of the correlations present in the measurements.
\end{abstract}


\maketitle

\section{Introduction}

\noindent Quantum teleportation (QT) \cite{Bennett1993} is one of the most astonishing procedures allowed by quantum systems. It consists of the transmission of quantum states from one system to another. \textit{Perfect teleportation} stands for a protocol with success probability one, while \textit{probabilistic teleportation} involves procedures in which transmission occurs with certain probability \cite{Agrawal2002}. Considering the existence of non-perfect procedures, it is possible to quantify the ability of a given protocol to transmit quantum states using different measures, being the \textit{average fidelity} the most common one \cite{Pirandola2015}. This functional quantifies the average overlap between the teleported states and the input states.
%\cite{Bosyk2014}.
One remarkable aspect of perfect QT is that it is impossible to achieve it by classical means \cite{Massar2005,Vidal1999}. Restricting us to teleport pure states, the maximal average fidelity reachable by classical procedures is $2/3$ \cite{Massar2005}.

In a \textit{standard teleportation} protocol, Alice performs a Bell-state measurement on the unknown state and one-half of the maximally entangled pair. Depending on the measurement outcome, Bob applies a suitable local unitary operation. When focusing, as usual, on teleportation of pure states, the resource state must be entangled to do better than a classical channel \cite{Popescu1994,Horodecki1999,Verstraete2003}. After Popescu's pioneering work \cite{Popescu1994},  about quantum teleportation with a mixed resource state, various authors have analyzed different aspects of this case in a variety of systems \cite{Verstraete2003,Guo2020,Roszak2015}. Mixed states shared by Alice and Bob represent more realistic conditions in which imperfections in the preparation of the proper state and the interaction with the environment can be taken into account. For non-maximally entangled resource states, the corresponding protocol has better fidelity than any classical communication protocol. Despite this, a non-zero amount of entanglement of the resource state is needed to have a protocol whose average fidelity is greater than the classical counterpart \cite{Verstraete2003}. While other forms of truly quantum teleportation certification have been recently discussed \cite{Cavalcanti2017}, the usual form consists in overcoming the maximal average fidelity that can be classically attained. 

On the other hand, it is important to recall that the classical protocol of teleportation of quantum states consists of a procedure that prepares states according to the optimal measurement process \cite{Massar2005,Vidal1999}. In the case of pure states, the optimal measurement process and its average fidelity over $N$ copies of the input state were determined by Massar and Popescu \cite{Massar2005}. In turn, Vidal \textit{et al.} determined the optimal measurement strategy for a two-state system prepared in a mixed state \cite{Vidal1999}. These results pave the way for studying the classical teleportation of mixed states. Concerning the quantum case, it is worth mentioning that while remote state preparation \cite{Bennett2001}, a variant of QT protocol, has been studied in several works for initial mixed states \cite{Berry2003, Xiang2005}, quantum teleportation has received much less attention when considering this kind of states. However, it is important to note that the input states of a QT protocol could  be quantum mechanically completely undefined, for instance when Alice's system is itself a member of an entangled pair and therefore has no well-defined properties \cite{Bennett1993}. Indeed, Bouwmeester \textit{et al.} experimentally demonstrated that the QT protocol is capable of teleporting an arbitrary (mixed and even an undefined) quantum state \cite{Bouwmeester1998}. Refs. \cite{Takei2005} and \cite{Jeong2007} are seminal examples of quantum teleportation of mixed (Gaussian) states within the continuous variable case. Additionally, initial mixed states and the role of their purity in determining the entanglement transfer were also considered within partial teleportation scenario \cite{Lee2000}.  

In this work, we extensively study the maximal average fidelity of teleportation when the protocol is intended to teleport mixed states. A question that naturally arises is: can separable states provide better-than-classical average fidelity of teleportation when this set of states is employed as the protocol resource? In this article,  we show that the standard quantum teleportation protocol with a \textit{separable state} as a resource overcomes the classical average fidelity if input mixed states are considered. This leads us to another question: are the quantum discord-like correlations present in the resource responsible for this quantum advantage? The answer is no. We show that even using classical-quantum states as a resource it is possible to overcome the average classical fidelity. To further understand these results we study how the maximal average fidelity depends on the correlations in Alice's measurements and its relation to the correlations of the resource state.

The article is organized as follow. In Sec. \ref{sec:QuantumTel} we briefly review the teleportation protocol and give the conditional states and their probabilities after Alice performs her measurement. In Sec. \ref{sec:average fidelity} we develop three key elements of our work: the average fidelity of a general protocol, the distributions of initial states, and Bob’s optimal operations. The classical protocol of teleportation, based on the seminal article of Vidal et al. \cite{Vidal1999}, is developed in Sec. \ref{sec:classicalprotocol}. In Sec. \ref{sec:MaximalfidelityofQT} we express the maximal average fidelity of the standard quantum teleportation protocol for an arbitrary isotropic distribution of initial states as a function of the \textit{fully entangled fraction}, see Sec. \ref{sec:satandardQT}. The particular cases of Werner and classical-quantum states as resources of the protocol are discussed in Sec \ref{sec:WandCQresourcesStandardQT}. In Sec. \ref{sec:changing measurement} we extend the previous result to a monoparametric family of measurements connecting the Bell measurement and the computational basis. In both cases, we fully analized the maximal average fidelity for  arbitrary mixed and quasi-pure states as input states of the protocol. Finally, a summary and conclusions are given in Section \ref{sec:conclusions}. For the sake of completeness we provide the explicit mathematical background related to Sec. \ref{sec:changing measurement} in the Appendix.

%Appendix B contains additional results considering input states with a fixed degree of purity. 

\section{Quantum teleportation protocol}\label{sec:QuantumTel}

\noindent Let us consider three qubits $A$, $B$, and $C$. The quantum teleportation protocol is carried out by two entities, Alice and Bob. The former has access to qubits $A$ and $B$, and the latter to systems $B$ and $C$. The ultimate goal of the protocol is to transmit the state of the system $A$ (unknown by Alice and Bob) to the qubit $C$, being
\begin{align}\label{eq:statetoteleport}
\rho_A=\frac{1}{2}\left(\mathbbm{1}_A+\vect{t}\cdot\vect{\sigma}\right),
\end{align}
with $\vect{t}\cdot\vect{\sigma}=\sum_j t^j \sigma_j$, $\{\sigma_i\}_{i=1}^3$ the set of Pauli matrices and $\vect{t}\in \mathbb{R}^3$ a column vector such that $\left|\vect{t}\right|\leq 1$. Additionally, the joint state $\rho$ of qubits $B$ and $C$ is commonly referred to as the \textit{resource state}. In order to carry out the teleportation of $\rho_A$, Alice measures $\mathcal{M}=\{M_i^{AB}\}$ over the two-qubit subsystem $AB$. Let us concentrate in arbitrary von Neumann measurements. This measurement affects the state of qubit $C$ according to
\begin{align}\label{eq:conditionalstates}
\rho_{C|i}&=\frac{\TrP{AB}{M^{AB}_i\otimes\mathbbm{1}_C \rho_A \otimes \rho}}{p_i}, \\
p_i&=\Trr{M^{AB}_i\otimes\mathbbm{1}_C \rho_A \otimes \rho}\label{eq:probcondstates},
\end{align}
being $p_i$ the probability of the outcome $i$ in Alice's measurement. Finally, Alice (classically) communicates  the outcome to Bob, who applies a suitable unitary operation over $\rho_{C|i}$ to reconstruct the input state \cite{Bennett1993,Horodecki1996}. 

The most common example of perfect teleportation takes place when the measurement $\mathcal{M}$ stands for the Bell basis $\{M^{AB}_i\}_{i=1}^4$
\begin{align}
M^{AB}_1&=\ket{\psi^+}\bra{\psi^+}, \label{eq:bell1} \\
M^{AB}_2&=\ket{\psi^-}\bra{\psi^-}, \label{eq:bell2}\\
M^{AB}_3&=\ket{\phi^+}\bra{\phi^+}, \label{eq:bell3}\\
M^{AB}_4&=\ket{\phi^-}\bra{\phi^-}, \label{eq:bell4} 
\end{align}
and the resource $\rho$ belongs to the previous set. Alternatively, if a Werner state is used as a resource,
\begin{align}\label{eq:WernerState}
\rho=pM^{BC}_1+\frac{(1-p)}{4}\mathbbm{1}_{B}\otimes\mathbbm{1}_C,
\end{align}
with $p\in [0,1]$, we have an example of probabilistic teleportation.  At the end of the protocol, after the optimal operations of Bob in qubit $C$, we obtain the following teleported state
\begin{align}\label{eq:standardoutomeWerner}
U_i\rho_{C|i}U_i^\dagger=\frac{1}{2}\left(\mathbbm{1}_C+p\vect{t}\cdot\vect{\sigma}\right)=\rho^{out}_C.
\end{align}
Let us go beyond the standard protocol considering $\mathcal{M}=\{M_i\}_{i=1}^4$ an arbitrary von Neumann measurement over $A$ and $B$, and $\rho$ any bipartite resource state corresponding to the systems $B$ and $C$. Then, each operator can be written as follows
\begin{align}\label{eq:measurementsvonNeu}
M_i&= \frac{1}{4}\left(\mathbbm{1}_{AB} + \sum_{k=1}^{3} n_i^k \sigma_k\otimes\mathbbm{1} + \sum_{l=1}^{3}m_i^l \mathbbm{1}\otimes\sigma_l + \sum_{k, l=1}^{3}C_i^{k l} \sigma_k\otimes\sigma_l\right), \\\label{eq:resourcestate}
\rho&= \rho_B\otimes\rho_C +\frac{1}{4}\sum_{k,l=1}^{3}T^{k l} \sigma_k\otimes\sigma_l,
\end{align}
being $\rho_B=\frac{1}{2}(\mathbbm{1}_B+\sum_k t_B^k \sigma_k)$ and 
$\rho_C=\frac{1}{2}(\mathbbm{1}_C + \sum_l t_C^l\sigma_l)$. In addition, if $C^{k l}=\Trr{\rho \sigma_k \otimes \sigma_l}$, it holds 
\begin{align}\label{eq:TandC}
T^{k l}=C^{k l}-t_B^k t_C^l.
\end{align}
Doing the required calculations, Eqs. \eqref{eq:conditionalstates} and \eqref{eq:probcondstates} result
\begin{align} \label{ec:conditionalstate2}
\rho_{C|i}&=\rho_C + \frac{1}{2}\sum_l \frac{\sum_k(m_i^k+\sum_j t^{j}C_i^{j k})T^{k l}}{4p_i}\sigma_l, \\
p_i&=\frac{1}{4}(1+\sum_k t^k n_i^k + \sum_l t_B^l m_i^l +\sum_{k l} t^k t_B^l C_i^{k l}).\label{eq:conditionalstateProb}
\end{align}
Thus, the Bloch vector corresponding to the conditional state $\rho_{C|i}$ can be written as
\begin{align}\label{eq:tconditionalstate}
\vect{t}_{C|i}=\frac{1}{4 p_i}\left[4 p_i \vect{t}_C + T^\intercal(\vect{m}_i +C_i^\intercal \vect{t})\right],
\end{align}
where the elements of matrices $T$ and $C_i$ are respectively $T^{k l}$ and $C_i^{k l}$.

\section{Average Fidelity}\label{sec:average fidelity}
\noindent Let us briefly review the most important ingredients for the computation of the average fidelity. If we consider a protocol that produces the output states $\rho^{out}_i$ with probability $p_i$, with $i\in [1,n]\subset \mathbb{N}$, for a given input state $\rho_A$, cf. Eq. \eqref{eq:statetoteleport}, we can quantify the score of the protocol by means of \cite{Massar2005,Vidal1999}
\begin{align}\label{eq:AverageFidelity1}
\mathcal{Q}(\rho_A)=\sum_{i=1}^n p_i F(\rho_A,\rho^{out}_i),
\end{align} 
being $F(\cdot,\cdot)$ the quantum fidelity, i.e.
\begin{align}
F(\rho_A,\rho^{out}_i)=\textrm{Tr}\left[\sqrt{\sqrt{\rho_A}\rho^{out}_i\sqrt{\rho_A}}\right]^2.
\end{align}
The ability of a given protocol to teleport states is quantified by the average of $\mathcal{Q}(\rho_A)$ over the input states. Let $f(\vect{t})$, $\vect{t}\in \mathcal{B}$, be the \textit{a priori} distribution of the input states in the unit ball $\mathcal{B}\subset\mathbb{R}^3$. 
The average fidelity results \cite{Vidal1999}
\begin{align}\label{eq:AverageFidelity F(t)}
\overline{F}=\int_{\mathcal{B}} \diff ^3\vect{t} f(\vect{t}) \mathcal{Q}(\rho_A),
\end{align}
being $\vect{t}$ the Bloch vector of $\rho_A$. 

In this article, we shall concentrate on isotropic distributions given by
\begin{align}
f(\vect{t})&=f(|\vect{t}|)=f(t), & 4\pi\int_0^1 \diff t \ t^2 f(t)&=1.
\end{align}

\noindent The case of random pure states as inputs of the protocol is obtained by taking \cite{Vidal1999}
\begin{align}
f_1(t)=\frac{1}{4\pi t^2} \lim_{t_0\to 1} \delta(t-t_0), \ t_0<1, \label{eq:puredistributionf}
\end{align}
giving rise to the following average fidelity:
\begin{align}\label{eq:averagefidPURE}
\overline{F}_1=\frac{1}{4\pi}\int_S \diff^2 \vect{t} \mathcal{Q}(\rho_A)
\end{align}
being $S$ the surface of the Bloch sphere. Another distribution of interest for this work corresponds to the case of random mixed states with a fixed level of purity given by 
\begin{align}\label{eq:distributionfixedpurity}
f_x(t)=\frac{1}{4\pi t^2}  \lim_{t_0\to x} \delta(t-t_0).
\end{align}
In the general case, i.e. random mixed states as input, the average fidelity results
\begin{align}\label{eq:AverageFidelity}
\overline{F}=\frac{3}{4\pi}\int_{\mathcal{B}} \diff^3 \vect{t} \mathcal{Q}(\rho_A),
\end{align}
which can be straightforwardly obtained from Eq. \eqref{eq:AverageFidelity F(t)} taking $f_m(t)=\frac{3}{4\pi}$. Finally, we consider random input states belonging to a shell given by $$S_{ab}=\{\vect{t}\in B \ / \ |\vect{t}|\in[a,b]\subset[0,1]\}.$$
The corresponding distributions is 
\begin{align}\label{eq:shell}
f_{ab}(t)=\frac{3}{4\pi(b^3-a^3)}\Theta(t-a)\Theta(b-t),
\end{align}
with $\Theta(x)$ the Heaviside step function.

\subsection{Optimal operations and maximal fidelity}\label{sec:optimalop}

\noindent Let us assume that we have a protocol with conditional states given by $\rho_{C|i}$, as for example in Eq. \eqref{eq:conditionalstates}. According to the quantum teleportation protocol, see Sec. \ref{sec:QuantumTel}, Bob has to apply suitable transformations over $\rho_{C|i}$ in order to reconstruct the initial state $\rho_A$. Therefore, the output states are $\rho^{out}_i=U_i \rho_{C|i} U_i^\dagger$ and the average fidelity, Eq. \eqref{eq:AverageFidelity}, depends on these operations $U_i$. Transformations that lead to the highest average fidelity are known as optimal operations. For qubit systems the score of the protocol, Eq. \eqref{eq:AverageFidelity1}, results \cite{Vidal1999}
\begin{align}\label{eq:scoreQubits}
\mathcal{Q}(\rho_A)= \frac{1}{2}\left[1+\vect{t}\cdot\left(\sum_i p_i \vect{t}_i\right)+\sqrt{1-\vect{t}^2}\sum_i p_i\sqrt{1-\vect{t}_{i}^2}\right],
\end{align} 
being $\vect{t}_i$ the Bloch vector of $\tilde{\rho_i}$. The action of the unitary transformation $U_i$ over the state $\rho_{C|i}$ corresponds to a rotation $R_i$ over the Bloch vector $\vect{t}$,  then it holds $\vect{t}_i=R_i\vect{t}_{C|i}$ with $\vect{t}_{C|i}$ the Bloch vector of $\rho_{C|i}$, \cite{Horodecki1996}. Thus, Bob's operations only affect the quantity
\begin{align}
\vect{t}\cdot\left(\sum_i p_i \vect{t}_i\right),
\end{align} 
and that is why it is necessary to optimize only this term over the input states. For convenience, we define
\begin{align}\label{eq:overlapp}
\tilde{\mathcal{Q}}(\rho_A)=\frac{1}{2}\left[1+\vect{t}\cdot\left(\sum_i p_i \vect{t}_i\right)\right],
\end{align}
and, correspondingly,
\begin{align}\label{eq:averagetilde}
\tilde{\overline{F}}=\int_{\mathcal{B}} \diff^3\vect{t} f(\vect{t}) \tilde{\mathcal{Q}}(\rho_A).
\end{align}
It is important to note that for $|\vect{t}|^2=1$, the last term of Eq. \eqref{eq:scoreQubits} vanishes. Then, for pure states hold  $\mathcal{Q}(\rho_A)=\tilde{\mathcal{Q}}(\rho_A)$ (and thus $\overline{F}=\tilde{\overline{F}}$).

Following Ref. \cite{Gu2004}, the general teleportation protocol described in Sec. \ref{sec:QuantumTel} can be written as a channel acting on the input state $\rho_A$. Moreover, the \textit{fully entangled fraction}, a measure of quantum correlations in bipartite systems \cite{Bennett1996}, has a direct relationship to the fidelity of teleportation maximized under the actions of local unitary operations. Specifically, the maximal average fidelity for initial arbitrary pure states, Eq. \eqref{eq:averagefidPURE}, corresponding to the standard quantum teleportation (i.e. the Alice's measurement is the Bell basis, Eqs. \eqref{eq:bell1}-\eqref{eq:bell4}) is given by
\begin{align}\label{eq:fullyandfidelity}
\overline{F}_{\max,1}=\frac{2\mathcal{F}(\rho)+1}{3},
\end{align} 
being $\mathcal{F}(\rho)$ the fully entangled fraction 
\begin{align}
\mathcal{F}(\rho)=\max_{W} \bra{\Phi} (W\otimes\mathbbm{1}) \rho (W^\dagger\otimes\mathbbm{1}) \ket{\Phi},
\end{align}
where $W$ is a 2$\times$2 unitary operator and $\ket{\Phi}$ is one of the maximally entangled Bell states. A simpler expression arises when the resource state $\rho$ is a Bell diagonal state
\begin{align}\label{eq:BellDiagState}
\rho_{\textrm{Bell}}=\frac{1}{4}\left(\mathbbm{1}_4+\sum_{i=1}^3c_i \sigma_i\otimes\sigma_i\right).
\end{align}
In this case, the fully entangled fraction results \cite{Guo2020}
\begin{align}\label{eq:fullyentangledBellDiagonalStates}
\mathcal{F}(\rho_{\textrm{Bell}})=\max_i \{\lambda_i\},
\end{align}
with $\{\lambda_i\}$ the eigenvalues of $\rho_{\textrm{Bell}}$ \cite{bussandri2020revisiting}.

\section{Classical Teleportation Protocol\label{sec:classicalprotocol}}
\noindent We now proceed with a classical teleportation protocol and explicitly calculate the average fidelity corresponding to random pure and mixed input states. If we have a unique copy of the input state $\rho_A$, the most general classical procedure to transmit $\rho_A$ to another system $C$ is the one proposed in Ref. \cite{Vidal1999}. 

In the first step of the protocol, Alice performs a measurement given by the POVM $\mathcal{M}_A=\{M_i\}_i$ with $n\in\mathbb{N}$ outcomes on system $A$. The probability of result $i$ is given by $p_i=\Trr{M_i \rho_A}$. When Alice obtains the outcome $i$, Bob (who has access to $C$) prepares a state $\rho^{out}_i$ following a \textit{guessing strategy}. Now, without loss of generality, let us consider \cite{Vidal1999}
\begin{align}
M_i=c_i^2\rho_i,
\end{align}
being $\rho_i=\frac{1}{2}(\mathbbm{1}_A+\vect{s}_i\cdot \vect{\sigma})$ and $c_i\in\mathbb{R}$. Then, in order to have
\begin{align}
\sum_i M_{i}=\sum_{i} c_i^2 \rho_i=\mathbbm{1}_A,
\end{align}
it must hold that
\begin{align}
\sum_i c_i^2&=2, & \sum_i c_i^2 \vect{s}_i&=0,
\end{align}
and thus the probability of the outcome $i$ results,
\begin{align}
p_i=\frac{c_i^2}{2}(1+\vect{s}_i\cdot \vect{t}).
\end{align}
The score \eqref{eq:AverageFidelity1} for the classical protocol reads
\begin{align}
\mathcal{Q}_c(\rho_A)&=\frac{1}{4}\sum_i c_i^2(1+\vect{s}_i\cdot \vect{t})\left(1+\vect{t}\cdot\vect{r}_i+\sqrt{1-t^2}\sqrt{1-r_i^2}\right),
\end{align}
where $\vect{r}_i$ is the Bloch vector of $\rho^{out}_i$.

Through the maximization of the corresponding average fidelity, cf. Eq. \eqref{eq:AverageFidelity F(t)}, Vidal \textit{et al.} determine the optimal measuring strategy (i.e. choice of $\vect{s}_i$) and the best guessing strategy (i.e. election of $\vect{r}_i$) \cite{Vidal1999}. The former corresponds to rank-one operators: $s_i^2=1$; while the latter is given by
\begin{align}\label{eq:guessing}
\vect{r}_i=\frac{(1-4I_1)\vect{s}_i}{\sqrt{36I_{1/2}^2+(1-4I_1)^2s_i^2}},
\end{align}
being
\begin{align}
I_\alpha=4\pi\int_0^1 \diff t \ t^2 f(t) \left(\frac{1-t^2}{4}\right)^\alpha.
\end{align}
The average fidelity corresponding to the best guessing strategy, Eq. \eqref{eq:guessing}, is
\begin{align}\label{eq:bestguessingAF}
\overline{F}_c =\frac{1}{4}\sum_{i} c_i^2\left(1+\frac{1}{3}\sqrt{36I_{1/2}^2+(1-4I_1)^2 |\vect{s}_i|^2}\right), 
\end{align}
while taking the optimal measuring strategy (rank-one operators) one has
\begin{align}
\max \overline{F}_c =\frac{1}{2}\left(1+\frac{1}{3}\sqrt{36I_{1/2}^2+(1-4I_1)^2}\right). \label{eq:fidelidadmaximaclasica}
\end{align}

As it is already known \cite{Massar2005, Vidal1999}, by restricting to random pure input states, we have

\begin{align}
\max \overline{F}_c \ \textrm{(pure)} =\frac{2}{3},\label{eq:classicalaveragefidelityPURE}
\end{align}
while if we one considers random mixed states as inputs 
\begin{align}
\max \overline{F}_c \ \textrm{(mixed)} &= \frac{1}{160}\left(80+\sqrt{256+225\pi^2}\right) \nonumber \\
&\approx 0.811,  \label{eq:maximumm fidelity}
\end{align}
is obtained. On the other hand, in the case of random mixed states with a fixed level of purity, Eq. \eqref{eq:distributionfixedpurity}, we obtain the following maximal average fidelity,
\begin{align}\label{eq:classicalaveragefixedputiry}
\max \overline{F}_{c,x}=\frac{1}{2}\left(1+\sqrt{1-x^2+\frac{x^4}{9}}\right),
\end{align}
where $x\in [0,1]$ is the modulus of the Bloch vector of the input states.

\section{Maximal fidelity teleporting mixed states}\label{sec:MaximalfidelityofQT}

\noindent In this section we develop the necessary expressions to obtain the maximal fidelity of teleportation in the general protocol exposed in Sec. \ref{sec:QuantumTel}, given by Eqs. \eqref{eq:measurementsvonNeu} and \eqref{eq:resourcestate}. The resulting conditional states $\rho_{C|i}$ have Bloch vector $\vect{t}_{C|i}$ and probability $p_i$, see Eqs. \eqref{eq:tconditionalstate} and \eqref{eq:conditionalstateProb}, respectively. As we saw in Sec. \ref{sec:optimalop}, to maximize the average fidelity of a given qubit protocol we should just optimize the average of $\tilde{\mathcal{Q}}(\rho_A)$, Eq. \eqref{eq:overlapp}. In consequence, by using Eqs. \eqref{eq:TandC}, \eqref{eq:conditionalstateProb} and \eqref{eq:tconditionalstate}, it is easy to show the following equality:
\begin{align}
p_i \vect{t}_{C|i}=\frac{1}{4}\left[\vect{t}_C(1+\vect{t}\cdot \vect{n}_i)+C^\intercal \vect{m}_i+(C_i C)^\intercal \vect{t}\right].
\end{align}
Therefore, bearing in mind $\vect{t}_i=R_i\vect{t}_{C|i}$, we have
\begin{align}
\frac{1}{2}\vect{t}\cdot\left(\sum_i p_i \vect{t}_i\right)&=\frac{1}{8}\left\{ \vect{t}\cdot \sum_i R_i (\vect{t}_C+C^\intercal \vect{m}_i)+\vect{t}\cdot \left[\sum_i R_i (\vect{t}_C\vect{n}_i^\intercal + C^\intercal C_i^\intercal)\right]\vect{t}\right\} \nonumber\\
&=\frac{1}{8}\left( \vect{t}\cdot \vect{\omega}+\vect{t}\cdot A\vect{t}\right).
\end{align}
At this point, it is important to note that vector $\vect{\omega}=\sum_i R_i (\vect{t}_C+C^\intercal \vect{m}_i)$ and matrix $A=\sum_i R_i (\vect{t}_C\vect{n}_i^\intercal + C^\intercal C_i^\intercal)$ do not depend on vector $\vect{t}$. Considering an isotropic distribution $f(\vect{t})=f(t)$, the average of $\tilde{\mathcal{Q}}(\rho_A)$ holds
\begin{align}
\tilde{\overline{F}}=\int_{\mathcal{B}} \diff^3 \vect{t} f(t)\tilde{\mathcal{Q}}(\rho_A)=\frac{1}{2}+\int_0^1 \diff t f(t) t^2 \int \diff^2 \vect{\Omega} \frac{1}{8}(\vect{t}\cdot\vect{\omega}+\vect{t}\cdot A\vect{t}),
\end{align}
where $\diff^2 \vect{\Omega}$ is a differential area over the surface of the Bloch sphere. Setting $\vect{t}=t\vect{x}$, with $\vect{x}^2=1$, and using the identities \cite{Horodecki1996}
\begin{align}
\int_S \diff^2 \vect{x} \ \vect{x}\cdot A\vect{x}&=\frac{4\pi}{3}\Trr{A}, \\
\int_A \diff^2 \vect{x} \ \vect{x}\cdot \vect{\omega}&=0,
\end{align}
for the integration over the angular variables, we have 
\begin{align}
\tilde{\overline{F}}=\frac{1}{2}+\frac{1}{8}\int_0^1 \diff t f(t) t^4 \int_S \diff^2 \vect{x} (\vect{x}\cdot\vect{\omega}+\vect{x}\cdot A\vect{x})=\frac{1}{2}+\frac{1}{8}\frac{4\pi \alpha}{3}\Trr{A},
\end{align}
being $\alpha=\int_0^1 \diff t f(t)t^4$. On the other hand, given that $M_i$ is a pure state we have $\vect{n}_i=C_i \vect{m}_i$, see Ref. \cite{Aravind1996}, and
\begin{align}
\Trr{A}&=\Trr{\sum_i R_i (\vect{t}_C\vect{n}_i^\intercal + C^\intercal C_i^\intercal)}=\sum_i \Trr{C_i(\vect{m}_i \vect{t}_C^\intercal+C)R_i^{-1}}.
\end{align}
In the following sections we will maximize $\tilde{\overline{F}}$ (and therefore the complete average fidelity) over Bob's operations, parametrized by the rotations $R_i$, i.e. 
\begin{align}\label{eq:optimization}
\tilde{F}_{\textrm{max}}=\frac{1}{2}+\frac{1}{8}\frac{4 \pi \alpha}{3}\max_{R_i} \Trr{A},
\end{align}
for some particular measurements $\{M_i\}$ and different resource states $\rho$.

\subsection{Standard quantum teleportation of mixed states \label{sec:satandardQT}}

\noindent Let us consider now that the Alice's measurement operators $\{M'_i\}$ are the Bell states, Eqs. \eqref{eq:bell1}-\eqref{eq:bell4}. Thus, $\vect{n}'_i=\vect{m}'_i=\vect{0}$ and $C'_i$ are diagonal improper rotations (orthonormal matrices with determinant equal to $-1$), specifically:
\begin{align}
C'_1&=\textrm{diag}(-1,-1,-1), \label{eq:CmatrixBell1}\\
C'_2&=\textrm{diag}(-1,1,1), \label{eq:CmatrixBell2}\\
C'_3&=\textrm{diag}(1,-1,1), \label{eq:CmatrixBell3}\\
C'_4&=\textrm{diag}(1,1,-1). \label{eq:CmatrixBell4}
\end{align}
In this case, we have $\Trr{A}=\sum_i \Trr{C'_iCR_i^{-1}}=\sum_i\Trr{CR_i^{-1}C'_i}$. Taking into account that the purity of the initial states, captured by $\alpha=\int_0^1 \diff t f(t)t^4$, does not play any role in the maximum in Eq. \eqref{eq:optimization}, it is possible to calculate it by using the results in Sec. \ref{sec:optimalop}. Specifically, for initial pure states, $\alpha=1/(4\pi)$, and employing Eqs. \eqref{eq:fullyandfidelity} and \eqref{eq:optimization}, we have
\begin{align}
\max_{R_i} \Trr{A}=8\left(2\mathcal{F}(\rho)-\frac{1}{2}\right) \implies
\tilde{\overline{F}}_{\textrm{max}}=\frac{1}{2}+\frac{4\pi \alpha}{3}\left(2\mathcal{F}(\rho)-\frac{1}{2}\right).
\end{align}
Therefore, the complete maximal average fidelity, see Eqs. \eqref{eq:AverageFidelity F(t)} and \eqref{eq:scoreQubits}, for the standard quantum teleportation protocol, results
\begin{align}\label{eq:fidelitystandartQT}
\overline{F}_{\textrm{max}}=\frac{1}{2}\left[1+\frac{8\pi\alpha}{3}\left(2\mathcal{F}(\rho)-\frac{1}{2}\right)+\int_{\mathcal{B}} \diff^3 \vect{t} f(t)\sqrt{1-\vect{t}^2}\sum_i p_i \sqrt{1-\vect{t}_{C|i}^2}\right],
\end{align}
where 	
\begin{align}
\vect{t}_{C|i}&=\frac{\vect{t}_C + (C_i C)^\intercal\vect{t}}{1+\vect{t}\cdot (C_i \vect{t}_B)},\\
p_i&=\frac{1}{4}[1+\vect{t}\cdot (C_i \vect{t}_B)].
\end{align}

\subsubsection{Werner and classical-quantum states as protocol resources}\label{sec:WandCQresourcesStandardQT}

\noindent If the resource state is a Bell diagonal state $\rho_{\textrm{Bell}}$ holds $\vect{t}_B=\vect{t}_C=\vect{0}$ and $C=\textrm{diag}(c_1,c_2,c_3)$ with $(c_1,c_2,c_3)\in \mathbb{R}^3$ belonging to the tetrahedron with vertices, Ref. \cite{Lang2010}, $$\{(-1,-1,-1),(-1,1,1),(1,-1,1),(1,1,-1)\}.$$
In this case, the Bloch vector of the conditional state $\rho_{C|i}$, with probability $p_i=1/4$, can be written as $\vect{t}_{C|i}=C_i C\vect{t}$, therefore $\vect{t}_{C|i}^2=(C\vect{t})^2$. The maximal fidelity of teleportation results, see Sec. \ref{sec:optimalop},
\begin{align}\label{eq:fidelitystandartQTBD}
\overline{F}_{\textrm{max}}=\frac{1}{2}\left[1+\frac{8\pi\alpha}{3}\left(2\mathcal{F}(\rho_{\textrm{Bell}})-\frac{1}{2}\right)+\int_{\mathcal{B}} \diff^3 \vect{t} f(t)\sqrt{1-\vect{t}^2} \sqrt{1-(C\vect{t})^2}\right],
\end{align}
with $\mathcal{F}(\rho_{\textrm{Bell}})$ given by Eq. \eqref{eq:fullyentangledBellDiagonalStates}. 

Considering a Werner state as the resource of the protocol, Eq. \eqref{eq:WernerState}, we have to take $C=\textrm{diag}(p,-p,p)$ and $\max_i \{\lambda_i\}$=$(1+3p)/4$ in Eq. \eqref{eq:fidelitystandartQTBD}. The maximal fidelity of teleportation in this case holds:
\begin{align}
\overline{F}_{\textrm{max}}&=\frac{1}{2}\left(1+4\pi\alpha p +
4\pi\int_0^1 dt \ t^2 f(t) \sqrt{(1-t^2)(1-p^2t^2)}\right). \label{eq:lastterm}
\end{align}
For random pure input states the last term in Eq. \eqref{eq:lastterm} vanishes, and the second one becomes $p$, because of $\alpha=1/(4\pi)$ if $f(t)=f_1(t)$. Thus,
\begin{align}
\overline{F}_{\textrm{max},1}=\frac{1}{2}(1+p).\label{eq:fideliyWernerPure}
\end{align}	
Therefore, as it is widely known, for $p\leq 1/3$ the average fidelity of this protocol is less than or equal to the average fidelity corresponding to the classical protocol, cf. Eq. \eqref{eq:classicalaveragefidelityPURE}.
We have a different scenario if  we consider instead random mixed input states setting $f_m(t)=3/(4\pi)$. In this case the  average fidelity results,
\begin{align}\label{eq:fidelityWernerMixed}
\overline{F}_{\textrm{max}}=\frac{1}{2}+\frac{3}{10}p + \frac{3}{2}\int_0^1dt \ t^2\sqrt{(1-p^2t^2)(1-t^2)}.
\end{align}
The integral in the last term can be written as $$\int_0^1dt \ t^2\sqrt{(1-p^2t^2)(1-t^2)}=\frac{2 \left(p^4-p^2+1\right) E\left(p^2\right)-\left(p^4-3 p^2+2\right) K\left(p^2\right)}{15 p^4},$$ for $p\in(0,1)$, being $E(m)$ and $K(m)$ the complete elliptic integral and the complete elliptic integral of the first kind, respectively. For $p=1$ it takes the value $2/15$ while if $p=0$ the integral turns out to be $\pi/16$. It is important to note a connection with the classical protocol. The case $p=0$ corresponds to take the product state $\frac{1}{2}\mathbbm{1}_B\otimes\frac{1}{2}\mathbbm{1}_C$ (i.e. no correlations of any kind between qubits $B$ and $C$) as a resource of the standard protocol, leading to $$\overline{F}_{\textrm{max}}=\frac{1}{2}+\frac{3}{2}\int_0^1dt \ t^2\sqrt{1-t^2}=\frac{1}{2}+\frac{3\pi}{32}.$$
This value coincides with the maximal average fidelity of the classical protocol \cite{Vidal1999} by performing no measurements, that is, Eq. \eqref{eq:bestguessingAF} with $|\vect{s}_i|=0$.

\begin{figure}[h]
	\centering
	\includegraphics[width=.5\textheight]{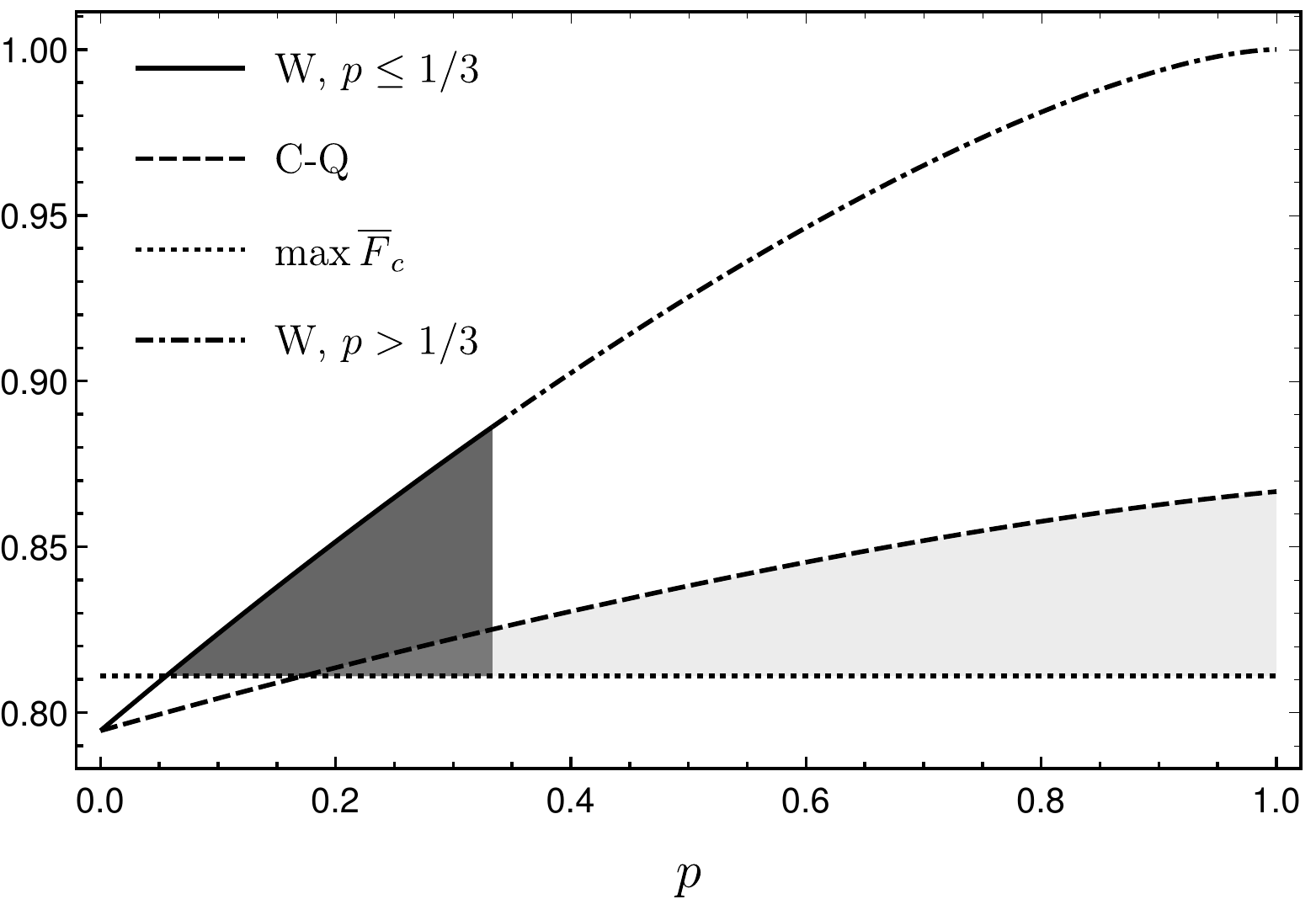}
	\caption{Maximal average fidelity for a Werner state ("W" curve, cf. Eq. \eqref{eq:fidelityWernerMixed}) and a classical-quantum state ("C-Q" curve, cf. Eq. \eqref{eq:fidelitystandartQTBDCQ}) as the resources of the standard teleportation protocol, considering random mixed input states as a function of $p\in [0,1]$. The horizontal dotted lines is the maximal average fidelity of the classical protocol (see Eq. \eqref{eq:maximumm fidelity}). All displayed quantities are dimensionless.}
	\label{fig:1}
\end{figure}

In Fig. \ref{fig:1} we plot the behavior of $\overline{F}_{\max}$ as a function of $p\in[0,1]$, showing that, when mixed input states are considered, there exist separable Werner states ($p\leq 1/3$) as resource of the standard QT protocol that lead to values of $\overline{F}_{\max}$ greater than the maximal average fidelity corresponding to the classical protocol $\max \overline{F}_c$, cf. Eq. \eqref{eq:maximumm fidelity}.

\begin{figure}[h]
	\centering
	\includegraphics[width=.5\textheight]{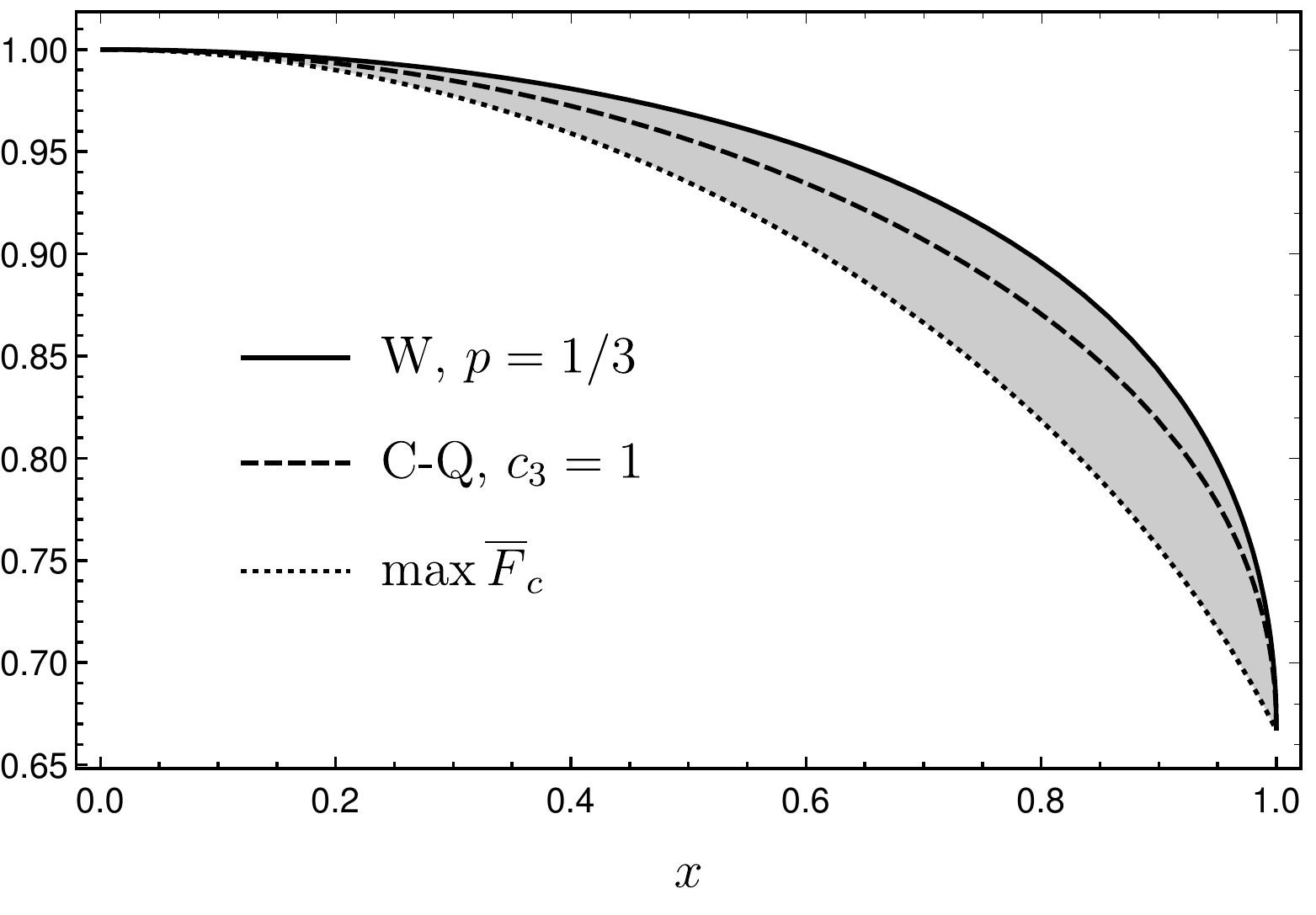}
	\caption{Maximal average fidelity considering random mixed input states with a fixed level of purity, cf. distribution given by Eq. \eqref{eq:distributionfixedpurity}, for two resource states: 1) Werner state ($\overline{F}_{\max,x}$ given by Eq. \eqref{eq:fixedpurityAF}) with $p=1/3$ ("W", full line) and 2) classical-quantum state ($\overline{F}_{\max,x}$ given by Eq. \eqref{eq:fidelitystandartQTBDCQ}) with $c=1$ ("C-Q", dashed line), as functions of $x\in [0,1]$. The dotted line correspond to the average fidelity of the classical protocol, Eq.\eqref{eq:classicalaveragefixedputiry}. All depicted quantities are dimensionless.}
	\label{fig:2}
\end{figure}

Additionally, we also consider an isotropic distribution corresponding to random mixed input states with a fixed purity (see Eq. \eqref{eq:distributionfixedpurity}). The maximal average fidelity for this distribution is
\begin{align}\label{eq:fixedpurityAF}
\overline{F}_{\max,x}=	\frac{1}{2}\left(1+px^2+\sqrt{(1-x^2)(1-p^2x^2)}\right).
\end{align}
In Fig. \ref{fig:2}, we show that the standard QT protocol with a Werner separable state (setting $p=1/3$) overcomes the corresponding classical average fidelity, Eq. \eqref{eq:classicalaveragefixedputiry}, for all $x\in(0,1)$.  The maximum difference $\Delta_\textrm{max}\approx 0.086$ is reached for $x \approx 0.904$. We also found  that lower values of $p$ lead to lower differences between $\overline{F}_{\max,x}$ and $\max \overline{F}_{c,x}$.

At this point, we could think that the quantum discord-type correlations existent in the Werner separable state are the key feature of the protocol in order to have maximal fidelity of teleportation greater than that corresponding to the classical protocol. However, we will see that even in the case of using classical-quantum states as protocol resources (zero discord-like correlations) we can overcome the classical average fidelity. Let us analyse this case by taking a simple set of these states: Bell diagonal classical-quantum states \cite{Lang2010}. Correspondingly, we have to choose $c_i=c$, $c_j=0$, $c_k=0$ with $(i,j,k)$ any permutation of $(1,2,3)$ and $-1\leq c \leq 1$. As we can see from Eq. \eqref{eq:fidelitystandartQTBD},  for isotropic distributions any choice leads to the same $\overline{F}_{\max}$. Let us take $c_1=c_2=0$ and $c_3=c$. Thus the maximal fidelity of teleportation results 
\begin{align}\label{eq:fidelitystandartQTBDCQ}
\overline{F}_{\textrm{max}}=\frac{1}{2}\left[1+\frac{4\pi\alpha c}{3}+\int_{\mathcal{B}} \diff^3 \vect{t} f(t)\sqrt{1-\vect{t}^2} \sqrt{1-(c t_3)^2}\right],
\end{align}
being $t_3$ the third coordinate of vector $\vect{t}$. In Figures \ref{fig:1} and \ref{fig:2} we can see the behavior of expression \eqref{eq:fidelitystandartQTBDCQ} ("C-Q" dashed line). In Fig. \ref{fig:1}, we plot $\overline{F}_{\max}$ as a function of $c=p\in [0,1]$ ($\overline{F}_{\max}$ is a symmetric function of $c$ in this case therefore we discard the case $c\in[-1,0]$), for random mixed input states i.e. $f(t)=3/(4\pi)$, showing there exists a range of $c$ in which the maximal fidelity of the standard quantum teleportation protocol overcomes the classical one even when the resource state is a Bell diagonal classical-quantum state (no discord-type quantum correlations in the resource of the protocol). In Fig. \ref{fig:2}, we show a similar situation if the input states have fixed purity given by $x$, i.e. $f(t)=f_x(t)$ as in Eq. \eqref{eq:distributionfixedpurity}. We take the most correlated classical-quantum state ($c=1$) and plot the maximal fidelity as a function of $x$ showing this quantity overcomes the classical one. The maximum average fidelity obtained with a classical-quantum state as a resource of the protocol is lower than the corresponding average fidelity to the most correlated Werner state ($p=1/3$).

\subsection{Changing measurement: from Bell to computational basis}\label{sec:changing measurement}

\noindent At the beginning of this section, we present the expressions to be optimized to obtain the maximal average fidelity when the resource is an arbitrary mixed state $\rho$, the Alice's measurement $\mathcal{M}=\{M_i\}$ is a (von Neumann) general one and Bob's operations to reconstruct the input states are unitary transformations. In Sec. \ref{sec:satandardQT}, we treated the standard quantum teleportation protocol, i.e. we set $\mathcal{M}$ as a Bell measurement, obtaining that the resulting maximal average fidelity is given by a term involving the fully entangled fraction plus another one related to the purities of the initial and conditional states, see Eq. \eqref{eq:fidelitystandartQT}. 

Let us consider now a more general measurement. Following Agrawal seminal work, Ref. \cite{Agrawal2002}, we take the projection operators onto the next four states as the Alice's measurement:
\begin{align}
\ket{\phi^+_l}&=\frac{1}{\sqrt{1+|l|^2}}(\ket{00}+l\ket{11}), \label{eq:agrawalstate1} \\
\ket{\phi^{-}_l}&=\frac{1}{\sqrt{1+|l|^2}}(l^*\ket{00}-\ket{11}), \\
\ket{\psi^+_l}&=\frac{1}{\sqrt{1+|p|^2}}(\ket{01}+p\ket{10}), \\
\ket{\psi^{-}_l}&=\frac{1}{\sqrt{1+|p|^2}}(p^*\ket{01}-l\ket{10}), \label{eq:agrawalstate4}	
\end{align}
being $l$ and $p$ complex numbers. It is important to note that the Bell basis is recovered if $p=l=1$ and, if $p=l=0$, we have the computational basis. For our next purpose it is convenient to rewrite the basis in polar coordinates, $l=r_l e^{i\phi_l}$ and $p=r_p e^{i\phi_p}$. We include the details of the measurement operators, marginal Bloch vectors and covariance matrices in the Appendix.

Regarding measurement correlations, the entanglement quantified by the von Neumann entropy results
\begin{align}
E(\ket{\phi_l^\pm})&=\frac{r_l^2\log_2r_l^2+(1-r_l^2)\log_2(1+r_l^2)}{1+r_l^2},\\
E(\ket{\psi_p^\pm})&=\frac{r_p^2\log_2r_p^2+(1-r_p^2)\log_2(1+r_p^2)}{1+r_p^2}.
\end{align}
An interesting case arises when all measurement projectors have the same amount of entanglement, Ref. \cite{Agrawal2002}, setting for example $l=p^*$. On the other hand, taking into account that the measurement correlations does not depend on the phases $\phi_l$ and $\phi_p$, we take $l$ and $p$ as real numbers: $l=p=r_n\in\mathbb{R}_{\geq 0}$. Therefore, we can write $C_i=D_nC'_i$ being $C'_i$ the corresponding matrix to the Bell states, Eqs. \eqref{eq:CmatrixBell1}-\eqref{eq:CmatrixBell4}, and $D_n=\textrm{diag}(c_n,c_n,1)$, with $0\leq c_n=2r_n/(1+r_n^2)\leq 1$ (see Appendix). It is important to note that $c_n$ captures the entanglement behavior, i.e. $E$ is a monotonic function of $c_n$. Correspondingly, we can directly choose $c_n$ as parameter for the measurement. On the other hand, if we additionally consider that the resource state of the protocol is a Bell diagonal state $\rho_{\textrm{Bell}}$, see Eq. \eqref{eq:BellDiagState}, it is straightforward to obtain the maximal average fidelity corresponding to this set up. We need to optimize Eq. \eqref{eq:optimization} with 
\begin{align}
\Trr{A}=\sum_i \Trr{D_nC'_i C R_i^{-1}}=\sum_i \Trr{C'_i D_nC R_i^{-1}}.
\end{align} 
As we can see, the optimization problem reduces to that presented in Sec. \ref{sec:satandardQT} with a different resource $\tilde{\rho}_{\textrm{Bell}}$ given by another (well-defined) Bell diagonal state with coefficients $c_n c_1$, $c_n c_2$ and $c_3$. Therefore,
\begin{align}
\tilde{\overline{F}}_{\textrm{max}}=\frac{1}{2}+\frac{4\pi \alpha}{3}\left(2\mathcal{F}(\tilde{\rho}_{\textrm{Bell}})-\frac{1}{2}\right).
\end{align}
However, this result does not imply that the entire protocol reduces to an standard quantum teleportation one. The maximal average fidelity of this protocol results
\begin{align} \label{eq:avergafidelity Werner+Agrawal}
\overline{F}_{\textrm{max}}=\frac{1}{2}\left[1+\frac{8\pi \alpha}{3}\left(2\mathcal{F}(\tilde{\rho}_{\textrm{Bell}})-\frac{1}{2}\right)+\int_{\mathcal{B}} \diff^3 \vect{t} f(t)\sqrt{1-\vect{t}^2}\sum_i p_i \sqrt{1-\vect{t}_{C|i}^2}\right],
\end{align}
being
\begin{align}
\vect{t}_{C|i}&=\frac{1}{4 p_i}\left[C(\vect{m}_i +C_i \vect{t})\right], \\
p_i&=\frac{1}{4}(1+t_3n_i),
\end{align}
where $C=\textrm{diag}(c_1,c_2,c_3)$, and $\vect{n}_i$, $\vect{m}_i$ and $C_i$, given by Eq. \eqref{eq:agrawaldecomp}. In the following, we will analyse the consequences of applying different measurements by changing $c_n$, for Werner and classical quantum-states (belonging to the set of Bell diagonal states) as the resources, and three particular isotropic distributions $f(t)$. 

\begin{figure}[h]
	\centering
	\includegraphics[width=.4\textheight]{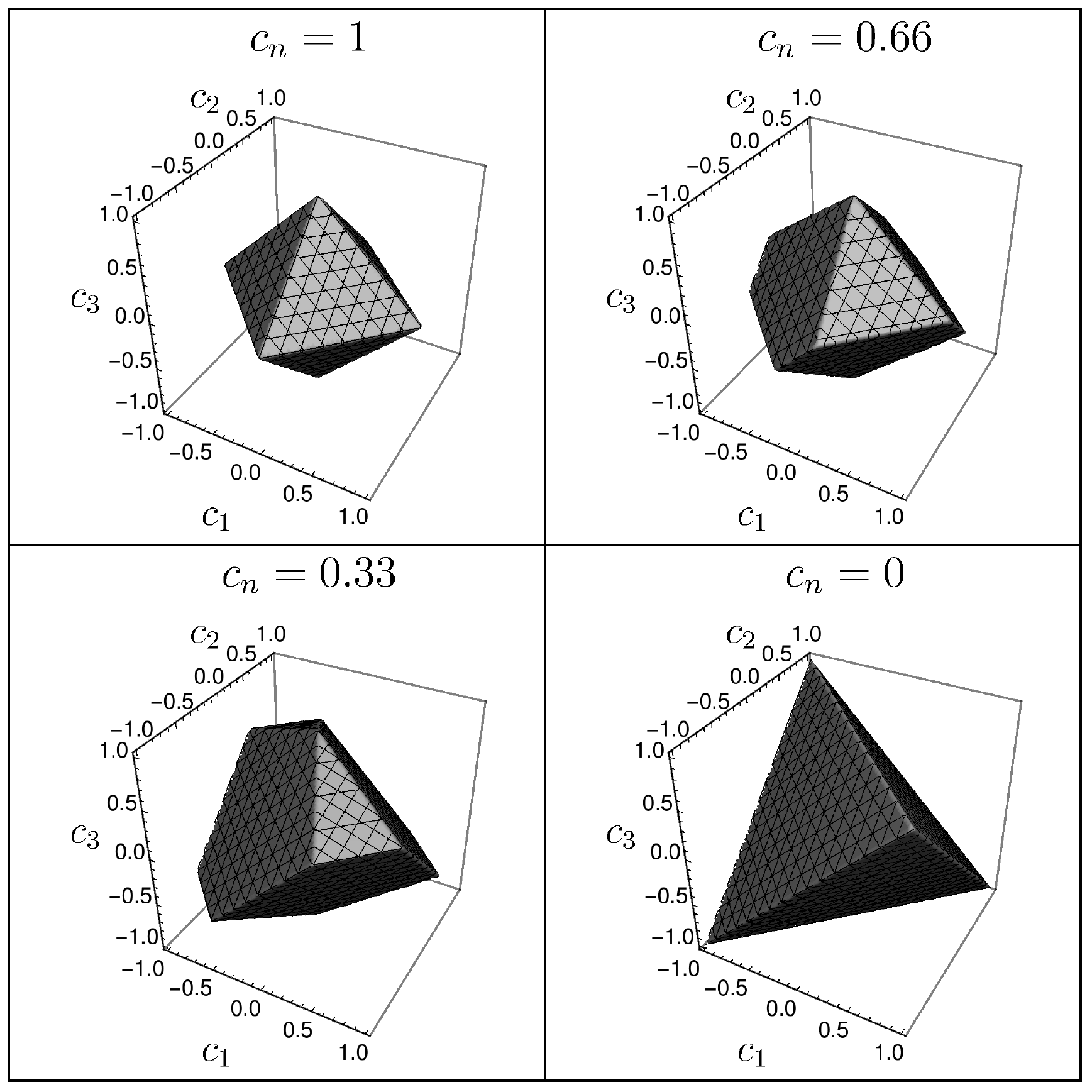}
	\caption{Volume defined by those $(c_1,c_2,c_3)$ implying maximal average fidelity of teleportation (pure input states), Eq. \eqref{eq:avergafidelity Werner+Agrawal}, lower than the maximal classical fidelity, $2/3$. $c_n$ stands for the measurement parameter. All quantities are dimensionless.}
	\label{fig:PureStatesAgrawal}
\end{figure}

\subsubsection{Pure states} 
\noindent Let us consider arbitrary initial pure states. Thus $\alpha=1/(4\pi)$ and
\begin{align}
\overline{F}_{\textrm{max}}=\frac{2\mathcal{F}(\tilde{\rho}_{\textrm{Bell}})+1}{3}.
\end{align}
In Fig. \ref{fig:PureStatesAgrawal} we plot the useless Bell diagonal states (namely, those states that leads to a maximal average fidelity of teleportation lower than the classical one) as we change the parameter $c_n$, going from the Bell measurement to the computational basis. As we can see, the volume of useless states increases as the measurement correlations tends to zero, case in which the volume coincides with the tetrahedron that defines the whole Bell diagonal states set. Therefore, no Bell diagonal state leads to a teleportation protocol with maximum average fidelity exceeding the classical one.

\begin{figure}[h]
	\centering
	\includegraphics[width=.4\textheight]{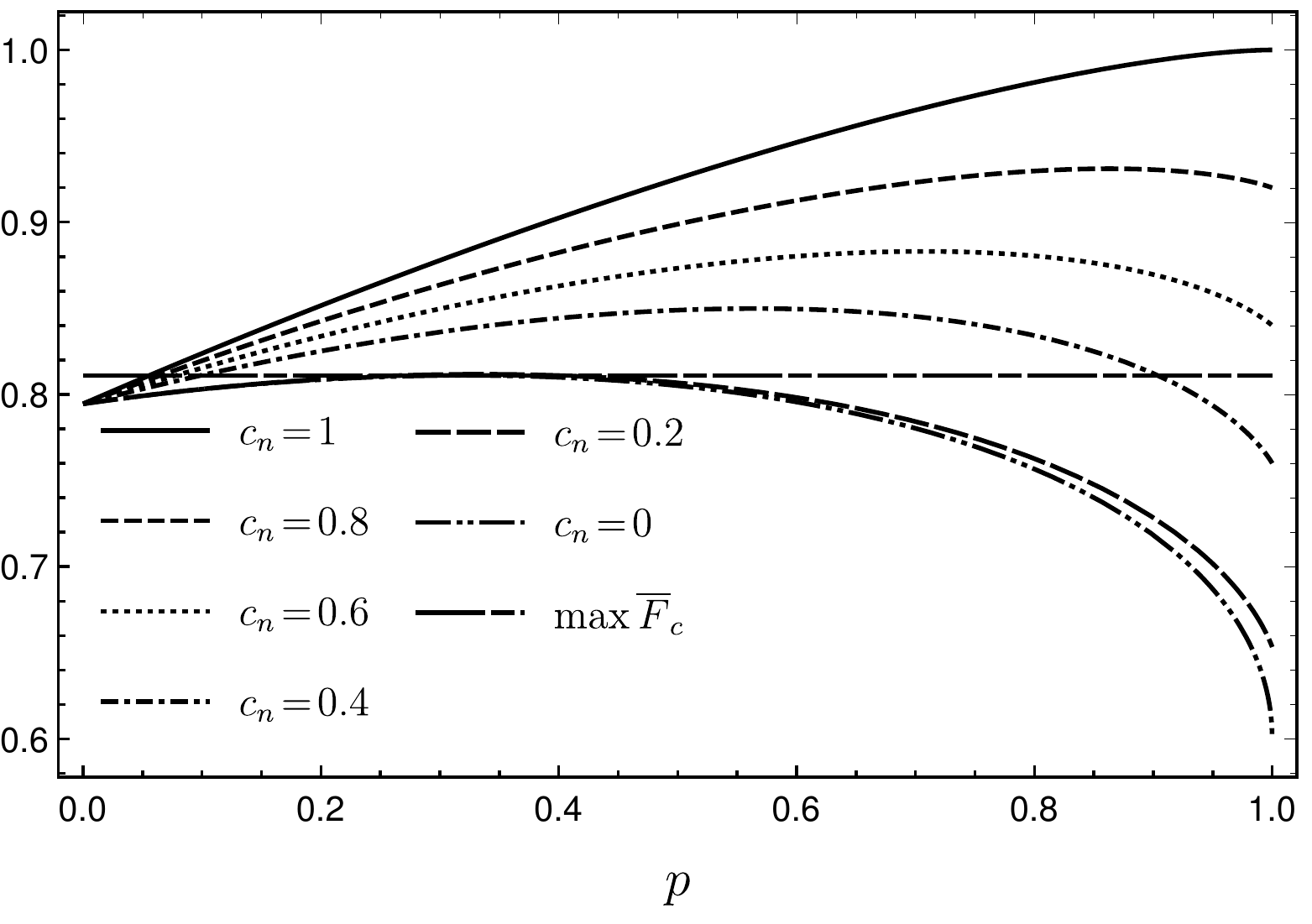}
	\caption{Maximal Average fidelity, Eq. \eqref{eq:avergafidelity Werner+Agrawal}, for a completely random distribution of input states and a Werner state, Eq. \eqref{eq:WernerState}, as the resource of the protocol. $c_n$ stands for the measurement parameter ($c_n=1$ is the Bell basis and $c_n=0$ is the computational basis). The horizontal line correspond to the maximal classical average fidelity, Eq. \eqref{eq:maximumm fidelity}.}
	\label{fig:FullMixWp}
\end{figure}

\subsubsection{Mixed input states} \label{sec:MixedInputStates}

\noindent Figure \ref{fig:FullMixWp} exhibits the behavior of Eq. \eqref{eq:avergafidelity Werner+Agrawal} in the case of completely mixed input states, i.e. $f(t)=3/(4\pi)$, being the resource of the protocol the Werner state \eqref{eq:WernerState} for which $C=\textrm{diag}(p,-p,p)$. The figure shows that the maximum average fidelity decreases for smaller values of the measurements quantum correlations. The curves from top to bottom correspond to $c_n=1$, $0.8$,  $0.6$, $0.4$, $0.2$, $0$. For uncorrelated measurements, $c_n=0$,the maximum average fidelity remains below the classical fidelity for all $p$.

\begin{figure}[h]
	\centering
	\includegraphics[width=.4\textheight]{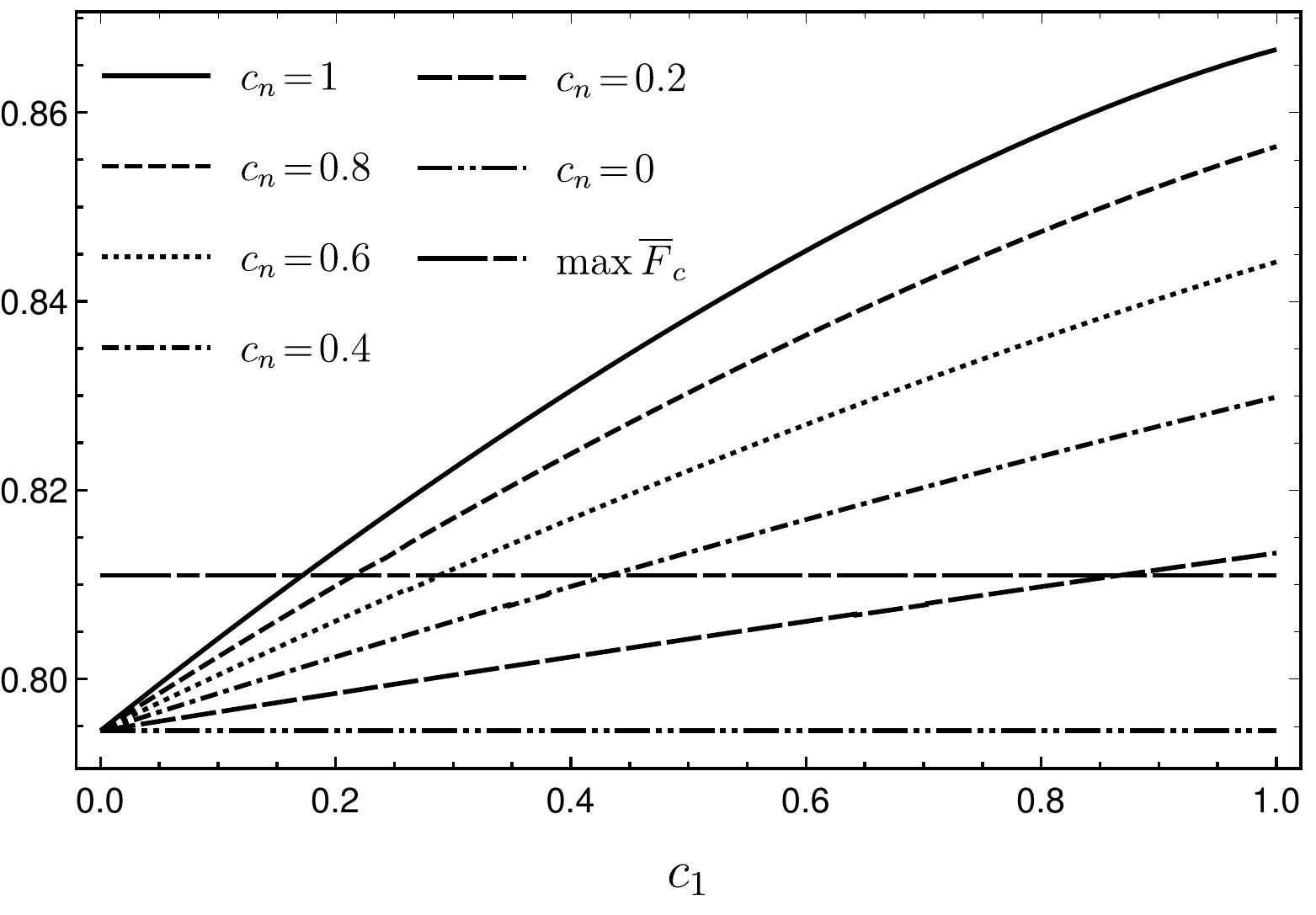}
	\caption{Maximal Average fidelity, Eq. \eqref{eq:avergafidelity Werner+Agrawal}, for a completely random distribution of input states and a classical-quantum (Bell diagonal) state given by $C=\textrm{diag}(c_1,0,0)$ as the resource of the protocol. $c_n$ stands for the measurement parameter ($c_n=1$ is the Bell basis and $c_n=0$ is the computational basis). The horizontal line correspond to the maximal classical average fidelity, Eq. \eqref{eq:maximumm fidelity}.}
	\label{fig:FullMixCQc1}
\end{figure}

\begin{figure}[h]
	\centering
	\includegraphics[width=.4\textheight]{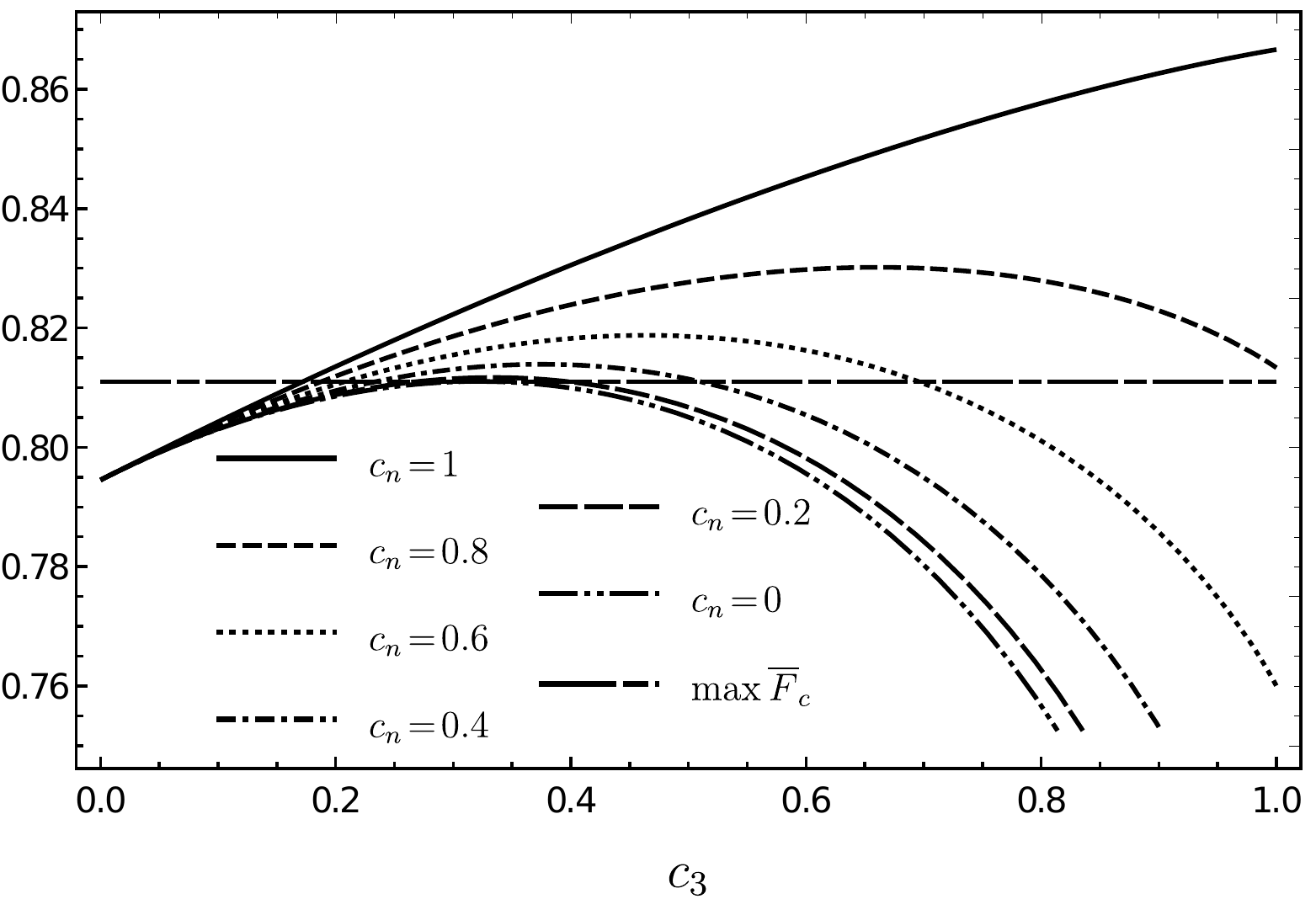}
	\caption{Maximal Average fidelity, Eq. \eqref{eq:avergafidelity Werner+Agrawal}, for a completely random distribution of input states and a classical-quantum (Bell diagonal) state given by $C=\textrm{diag}(0,0,c_3)$ as the resource of the protocol. $c_n$ stands for the measurement parameter ($c_n=1$ is the Bell basis and $c_n=0$ is the computational basis). The horizontal line correspond to the maximal classical average fidelity, Eq.  \eqref{eq:maximumm fidelity}.}
	\label{fig:FullMixCQc3}
\end{figure}

In Figs. \ref{fig:FullMixCQc1} and \ref{fig:FullMixCQc3} we include the behavior of the maximal fidelity for completely mixed input states but considering a classical-quantum resource. While for a protocol in which Alice measures the Bell basis the maximum average fidelity does not depend on the choice of state, see Eq. \eqref{eq:fidelitystandartQTBDCQ}, here different behaviors are observed depending on the choice of the classical-quantum Bell diagonal resource state. Fig. \ref{fig:FullMixCQc1} corresponds to the maximal average fidelity in which the resource is given by $C=\textrm{diag}(c_1,0,0)$ while Fig. \ref{fig:FullMixCQc3}, $C=\textrm{diag}(0,0,c_3)$. For state $(c_1, 0, 0)$ the maximum average fidelity is an increasing function of $c_1$ and also grows with $c_n$. For all $c_n\neq 0$ (in the figure) the maximum average fidelity exceeds the classical average fidelity when $c_1$ is sufficiently large. A relationship is observed between the amount of correlations in the resource state (determined by $c_1$) and the degree of correlations in the measurement base (given by $c_n$). 

In contrast, if the resource state is a Bell diagonal state with $C=\textrm{diag}(0,0,c_3)$, the relationship between its correlations and the amount of correlations in the measurement basis is more complex, see Fig. \ref{fig:FullMixCQc3}. In general, the maximal average fidelity is a concave function of $c_3$ for all $c_n$ being an increasing function only for the Bell basis $c_n=1$. For each intermediate value of $c_n$ there is a well-defined range of $c_3$ in which the average maximum fidelity exceeds the classical average fidelity. In all cases, there is a critical $c_3$ value, $c_3^*$, above which the maximum fidelity decreases and this critical value depends strongly on $c_n$. Curves corresponding to smaller $c_n$ values have associated smaller $c_3^*$ values. In all cases, the average maximum fidelity exceeds the classical fidelity by a minor difference for decreasing values of $c_n$ and does not exceed it in the case $c_n=0$.

%\change{See Appendix B for additional results regarding to initial states with fixed purity.}

\begin{figure}[h]
	\centering
	\includegraphics[width=.4\textheight]{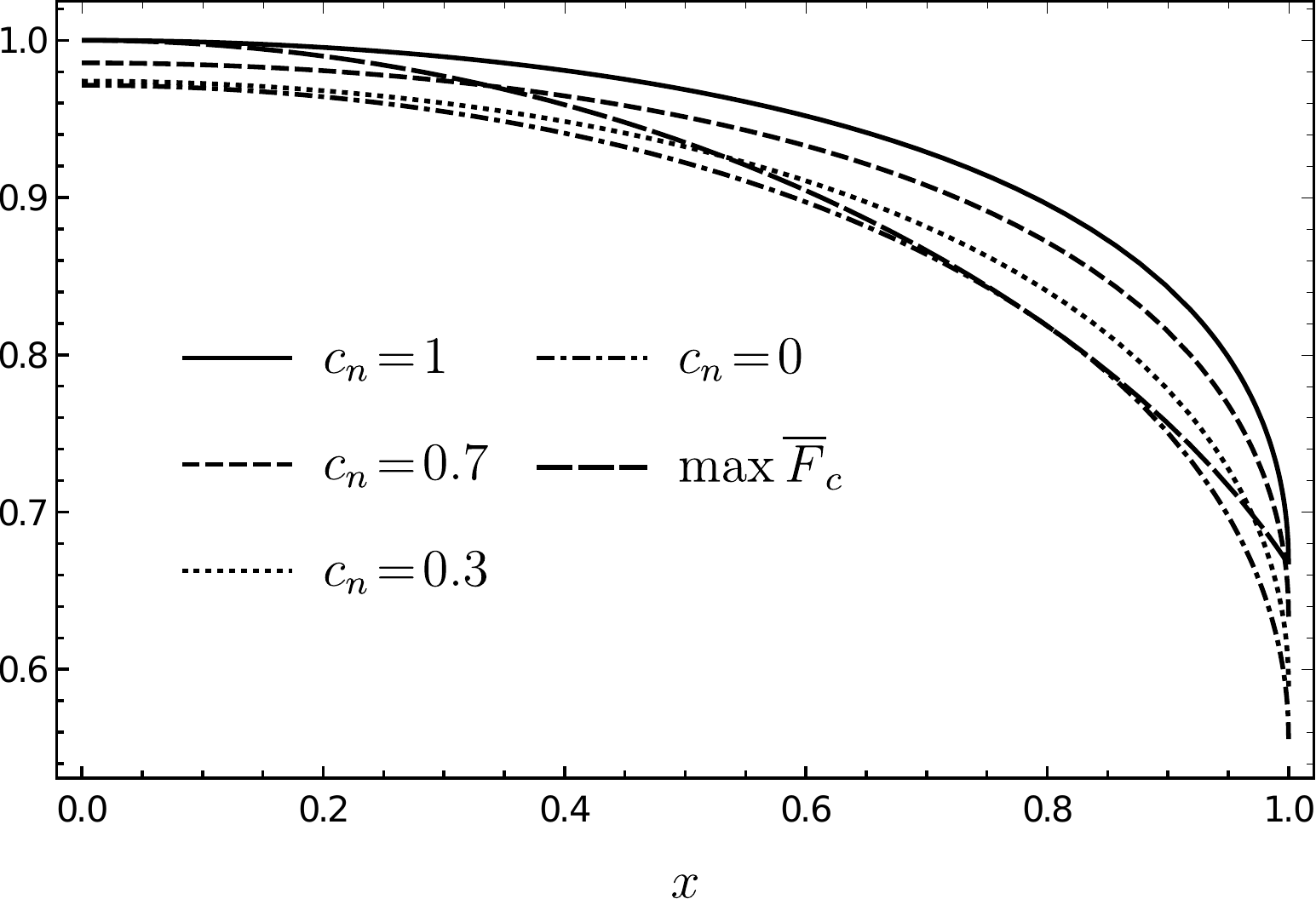}
	\caption{Maximal Average fidelity, Eq. \eqref{eq:avergafidelity Werner+Agrawal}, for a distribution $f_x$ given by Eq. \eqref{eq:distributionfixedpurity}, as a function of $x=|\vect{t}|$ being $\vect{t}$ the Bloch vector of the input states. The resource of the protocol is a Werner state, Eq. \eqref{eq:WernerState}, with $p=1/3$. $c_n$ stands for the measurement parameter ($c_n=1$ is the Bell basis and $c_n=0$ is the computational basis). $\max \overline{F}_c$ is the classical average fidelity, Eq. \eqref{eq:classicalaveragefixedputiry}.}
	\label{fig:FixedPurityWp}
\end{figure}

\begin{figure}[h]
	\centering
	\includegraphics[width=.4\textheight]{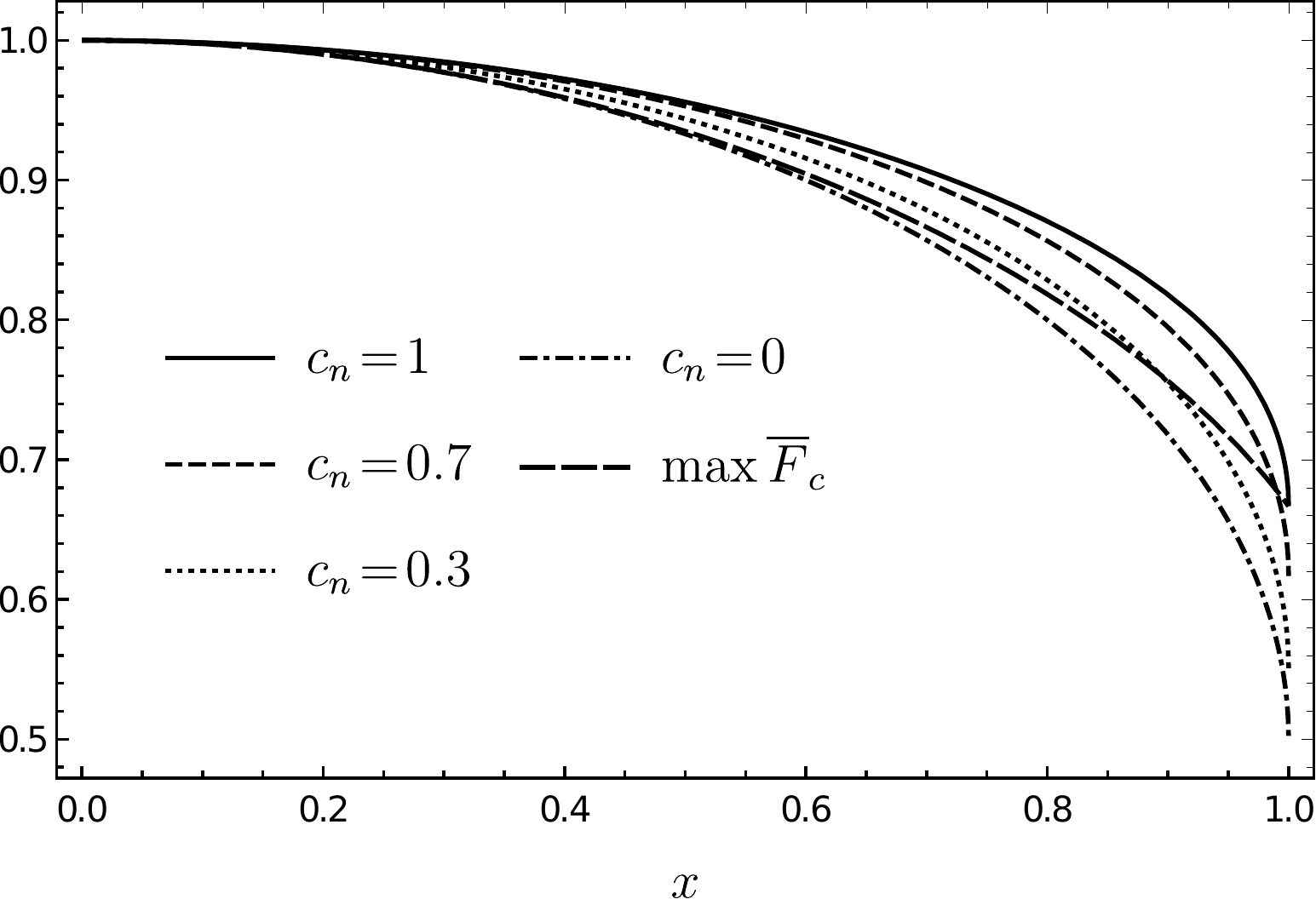}
	\caption{Maximal Average fidelity, Eq. \eqref{eq:avergafidelity Werner+Agrawal}, for a distribution $f_x$ given by Eq. \eqref{eq:distributionfixedpurity}, as a function of $x=|\vect{t}|$ being $\vect{t}$ the Bloch vector of the input states. The resource of the protocol is a classical-quantum (Bell diagonal) state, given by $C=\textrm{diag}(1,0,0)$. $c_n$ is the measurement parameter ($c_n=1$ is the Bell basis and $c_n=0$ is the computational basis). $\max \overline{F}_c$ stands for the classical average fidelity, Eq. \eqref{eq:classicalaveragefixedputiry}.}
	\label{fig:FixedPurityCQc1}
\end{figure}

\begin{figure}[h]
	\centering
	\includegraphics[width=.5\textheight]{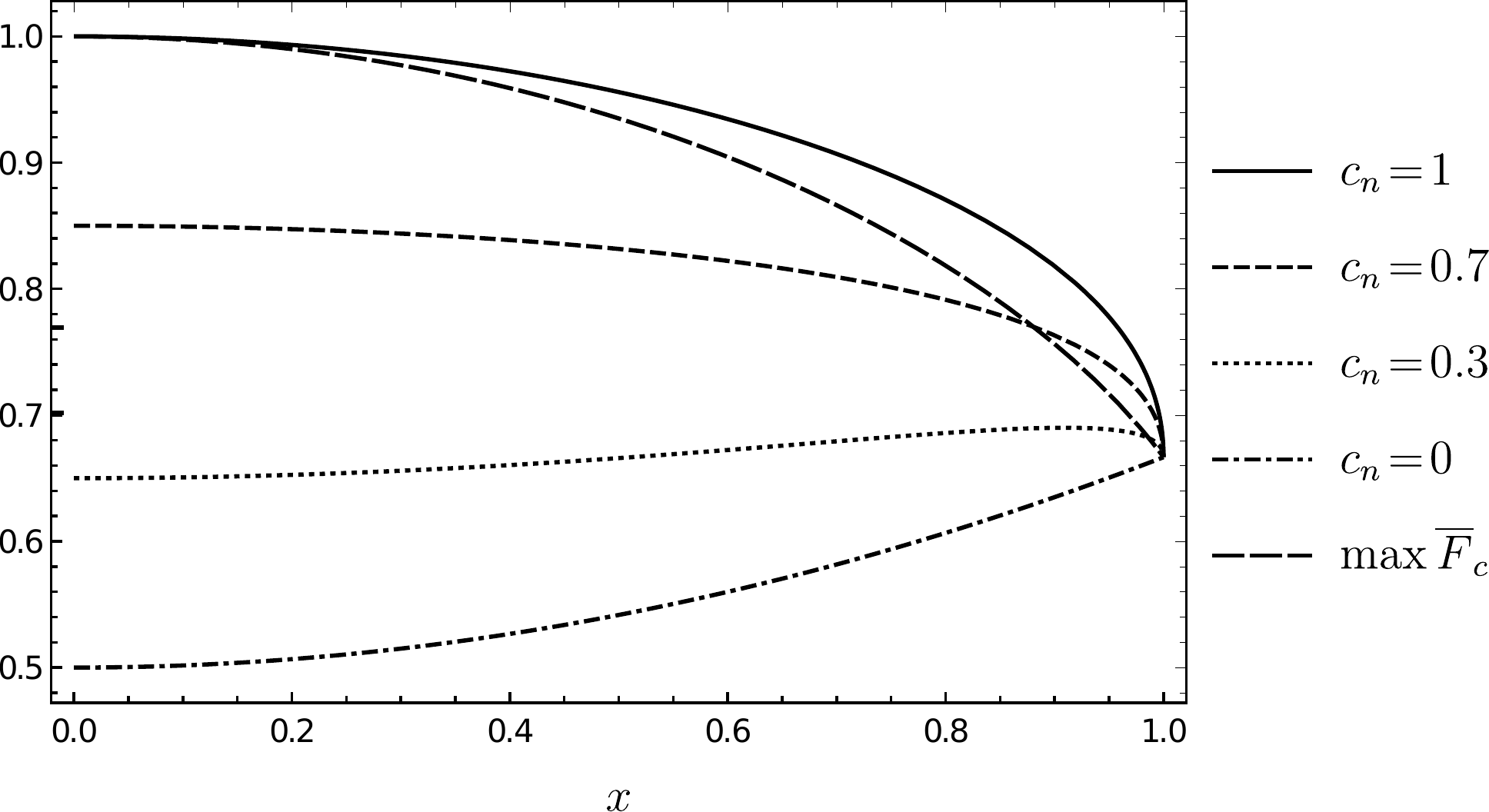}
	\caption{Maximal Average fidelity, Eq. \eqref{eq:avergafidelity Werner+Agrawal}, for a distribution $f_x$ given by Eq. \eqref{eq:distributionfixedpurity}, as a function of $x=|\vect{t}|$ being $\vect{t}$ the Bloch vector of the input states. The resource of the protocol is a classical-quantum (Bell diagonal) state, given by $C=\textrm{diag}(0,0,1)$. $c_n$ is the measurement parameter ($c_n=1$ is the Bell basis and $c_n=0$ is the computational basis). $\max \overline{F}_c$ stands for the classical average fidelity, Eq. \eqref{eq:classicalaveragefixedputiry}.}
	\label{fig:FixedPurityCQc3}
\end{figure}

Let us analyse the maximal average fidelity for an isotropic distribution of states with fixed purity, i.e. $f(t)$ given by Eq. \eqref{eq:distributionfixedpurity}. Figures \ref{fig:FixedPurityWp} displays the maximal  average fidelity as a function of the purity (given by $x$, the Bloch vector modulus corresponding to the initial inputs) for a teleportation protocol with the most correlated separable Werner state ($p=1/3$) as a resource and for different correlations of the measurement basis. The maximal average fidelity exceeds its classical counterpart for any purity value in the case of a standard teleportation protocol (Bell basis, $c_n=1$). A high enough degree of mixing in the input states is necessary for the maximal average fidelity to exceed the classical one for $c_n=0.7$ (this changes in the neighbourhood of $x=1$). Lower correlations in the measurement basis lead to maximal average fidelity below the classical fidelity. When the resource state corresponds to a classical-quantum (Bell diagonal) state, given by $C = \textrm{diag}(1, 0, 0)$, similar behavior is observed, see Fig. \ref{fig:FixedPurityCQc1}. In this case less correlated bases ($c_n=0.3$) also lead to average maximal fidelity that exceeds the classical fidelity even for states with a lower degree of mixing (higher purity). A different scenario is observed in Fig. \ref{fig:FixedPurityCQc3}, where a classical-quantum (Bell diagonal) state given by $C=\textrm{diag}(0,0,1)$ is considered as a resource. The classical average fidelity is outperformed by the corresponding quantum counterpart for highly correlated bases and input states with a high degree of purity. For less purity the maximal average fidelity falls well below the classical fidelity and a change in the convexity of the curve is also observed when $c_n \to 0$ (low correlated bases).

\begin{figure}[h]
	\centering
	\includegraphics[width=.55\textheight]{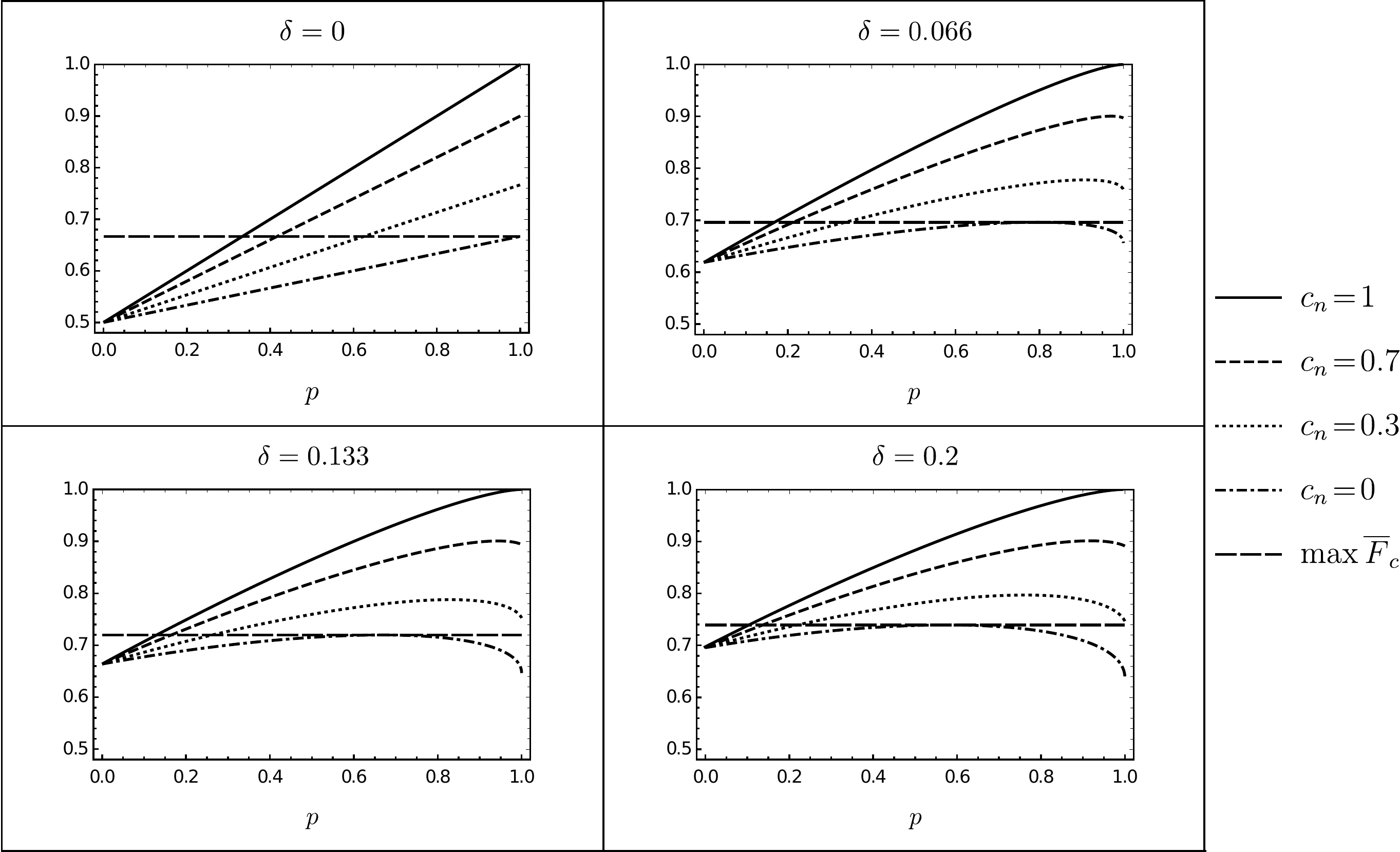}
	\caption{Maximal Average fidelity, Eq. \eqref{eq:avergafidelity Werner+Agrawal}, for a distribution $f_{ab}$, Eq. \eqref{eq:shell} with $b=1$ and $a=1-\delta$. The resource of the protocol is a Werner state, Eq. \eqref{eq:WernerState}. $c_n$ is the measurement parameter ($c_n=1$ is the Bell basis and $c_n=0$ is the computational basis). $\max \overline{F}_c$ stands for the classical average fidelity, Eq. \eqref{eq:fidelidadmaximaclasica}.}
	\label{fig:ShellWp}
\end{figure}
\begin{figure}[h]
	\centering
	\includegraphics[width=.55\textheight]{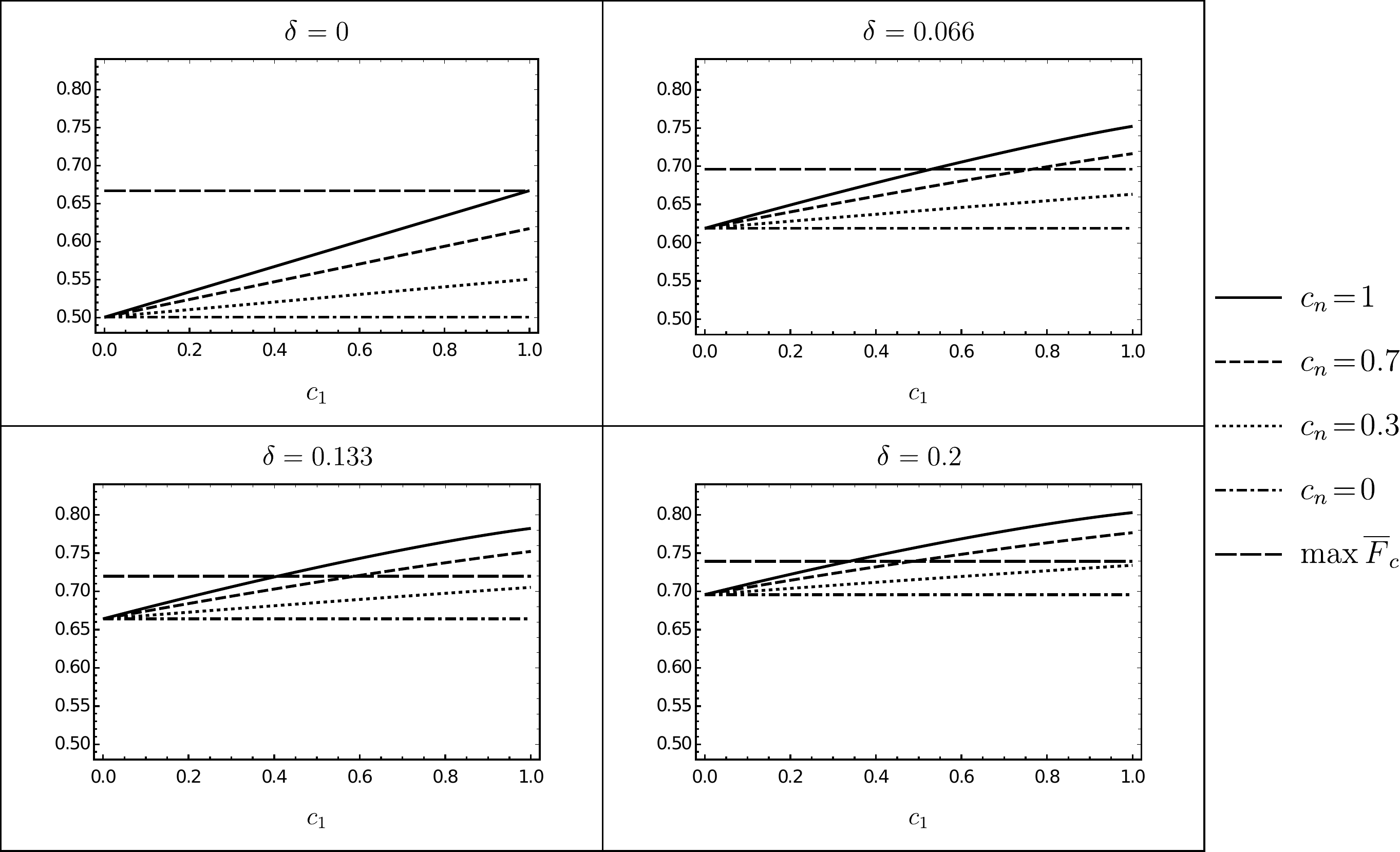}
	\caption{Maximal Average fidelity, Eq. \eqref{eq:avergafidelity Werner+Agrawal}, for a distribution $f_{ab}$, Eq. \eqref{eq:shell} with $b=1$ and $a=1-\delta$. The resource of the protocol is a classical-quantum (Bell diagonal) state, given by $C=\textrm{diag}(c_1,0,0)$. $c_n$ is the measurement parameter ($c_n=1$ is the Bell basis and $c_n=0$ is the computational basis). $\max \overline{F}_c$ stands for the classical average fidelity, Eq. \eqref{eq:fidelidadmaximaclasica}.}
	\label{fig:Shellc1}
\end{figure}

\begin{figure}[h]
	\centering
	\includegraphics[width=.55\textheight]{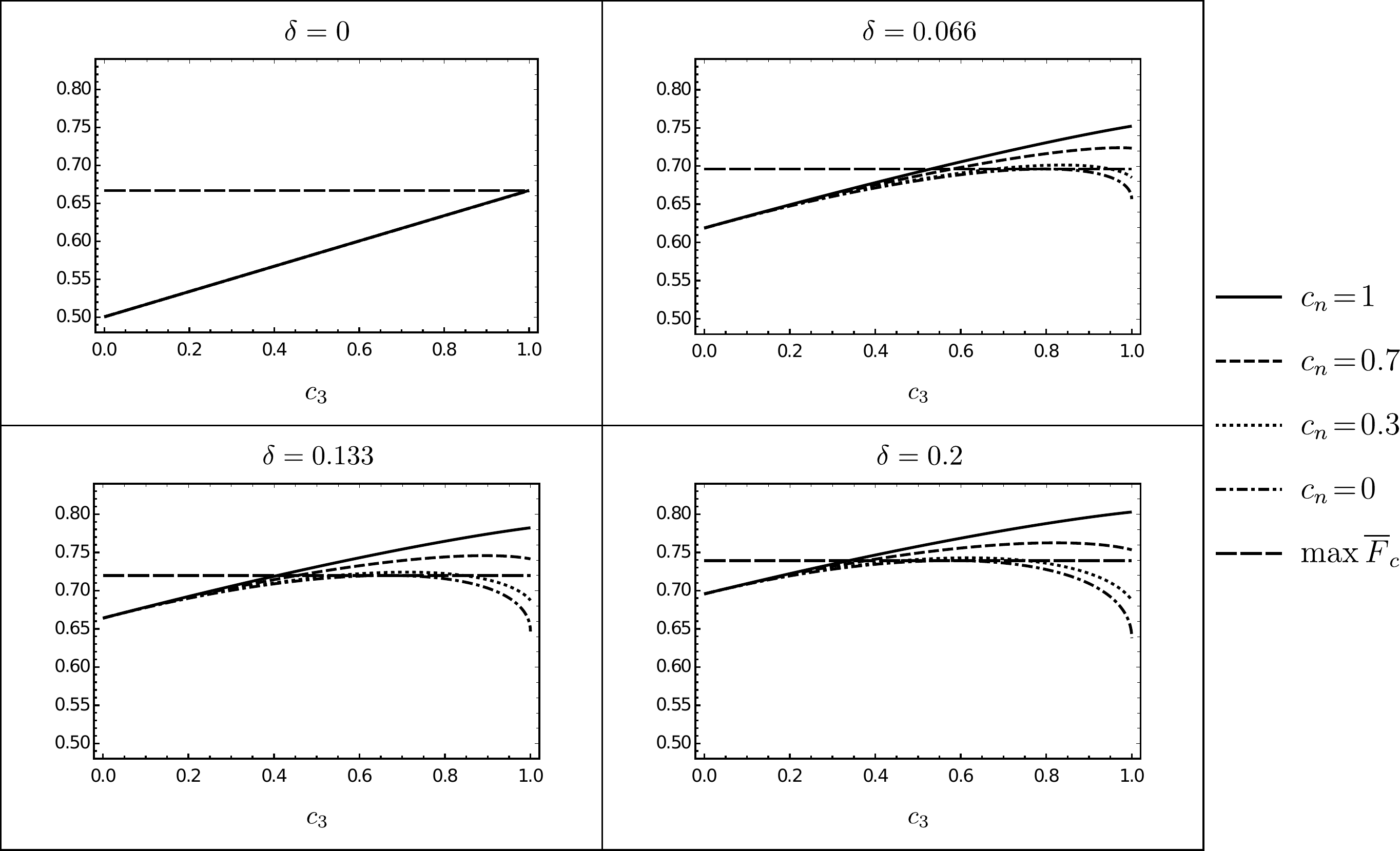}
	\caption{Maximal Average fidelity, Eq. \eqref{eq:avergafidelity Werner+Agrawal}, for a distribution $f_{ab}$, Eq. \eqref{eq:shell} with $b=1$ and $a=1-\delta$. The resource of the protocol is a classical-quantum (Bell diagonal) state, given by $C=\textrm{diag}(0,0,c_3)$. $c_n$ is the measurement parameter ($c_n=1$ is the Bell basis and $c_n=0$ is the computational basis). $\max \overline{F}_c$ stands for the classical average fidelity, Eq. \eqref{eq:fidelidadmaximaclasica}.}
	\label{fig:Shellc3}
\end{figure}

\subsubsection{Quasi-pure input states} 
\noindent Let us consider now quasi-pure input states distributed according to Eq. \eqref{eq:shell} with $b=1$ and $a=1-\delta$ being $\delta$ the shell width near the pure states. The maximal average fidelity corresponding to this distribution is plotted in Figs. \ref{fig:ShellWp} (for a Werner state resource with $C=\textrm{diag}(p,-p,p)$), \ref{fig:Shellc1} and \ref{fig:Shellc3} (for classical-quantum Bell diagonal states as the resources with $C=\textrm{diag}(c_1,0,0)$ and $C=\textrm{diag}(0,0,c_3)$, respectively). In all plots, the top-left figure corresponds to pure input states. As we can see, in this case the maximal fidelity of teleportation does not overcome the corresponding classical one if the resource belongs to the set of separable Bell diagonal states (that includes the classical-quantum states). However, if we increment in a small quantity the width $\delta$, allowing the inputs to have a small degree of mixture, we start to see how the maximal average fidelity can overcome the classical counterpart even when the resource is a separable state (Fig. \ref{fig:ShellWp}) or a classical-quantum state (Figs. \ref{fig:Shellc1} and \ref{fig:Shellc3}). Besides, we can observe from each plot that for greater values of $\delta$, the maximal average fidelity behavior tends to the corresponding one to mixed input states, see Sec. \ref{sec:MixedInputStates}, changing from a linear behavior to a concave function of correlations of the particular considered resource.

\section{Concluding remarks}\label{sec:conclusions}

\noindent Summarizing in this paper we compute the maximal average fidelity of a quantum teleportation protocol in which the states to be teleported form an isotropic distribution, Alice can perform any arbitrary measurement from a monoparametric family of von Neumann measurements, which includes the computational and Bell basis, and Bob applies unitary operations to reconstruct the input state. Two different scenarios are solved depending on the resource state of the protocol: 1) the resource state is arbitrary and Alice performs a Bell measurement, 2) the resource state is a Bell diagonal state and Alice performs some arbitrary measurement on the parametrized basis. For our calculations, we use the well-known relation between the maximal average fidelity and the fully entangled fraction for the standard protocol. We take advantage of the fact that only a part of the obtained expression for the average fidelity must be maximized (since the rest depends on the modulus of the Bloch vector of the input states and is not affected by the operations performed by Bob, see Sec. \ref{sec:optimalop}) and that the distribution of initial states is a multiplicative factor in the expression ($\alpha$ in the text) also not affected by Bob’s operations, see Eq. \eqref{eq:optimization}. It is important to recall that the latter results is also valid in the general case, namely, for arbitrary von Neumann measurements and resource states. 

We focus our calculations on different distributions of initial states $\rho_A$: 1) completely mixed input states, 2) states with a certain degree of purity given by $x$ (the Bloch vector modulus corresponding to the input states), and 3) mixed states with a degree of purity in a certain range, in particular, we study quasi pure states with purity in the range $x\in[1-\delta, 1]$.

Our results show that the standard quantum teleportation protocol can teleport arbitrary mixed states with higher average fidelity than the classical fidelity even in the case of using a separable (non-entangled) resource state. In particular, we show that the average teleportation fidelity outperforms the classical one when we use a separable (quantum correlated) Werner state. We also demonstrate that by using a maximally discordant separable Bell state ($p=1/3$) the maximal average fidelity outperforms its classical counterpart when teleporting states with a fixed degree of mixing $x\in(0,1)$, and that the maximum difference with classical fidelity occurs when teleporting quasi-pure states (i.e. with a degree of purity $x \approx 0.904$).

In order to answer whether it is the quantum discord-like correlations present in the resource states that make separable states useful for teleportation, we analyse the case of classical-quantum states (also diagonal in the Bell basis). We find that these states used as a resource in a standard teleportation protocol allow us to overcome classical fidelity when teleporting arbitrary mixed states and also in the case of mixed states with a fixed degree of purity.

Finally, we analyse the role played by the correlations present in the chosen measurement basis going out of the standard protocol and considering a different family of von Neumann measurements that allowed us to go from the Bell basis (standard protocol) to the computational basis (fully uncorrelated). We obtain the maximal average fidelity by reducing the maximization problem to that of the standard protocol previously solved. We show that when teleporting pure states the volume of non-useful Bell states (i.e. those states used as a resource leading to a protocol whose maximal average fidelity does not exceed the classical fidelity) increases as the correlations of the basis decrease until the whole tetrahedron representing the set of Bell diagonal states is reached \cite{Lang2010}. Namely, if Alice measures a projection onto the computational basis, no Bell diagonal state (regardless of its degree of entanglement) can generate a quantum protocol with maximal average fidelity higher than the classical one. The relationship between the entanglement of the state used as a resource and the entanglement of the basis on which Alice performs her measurements is analysed in detail in Ref. \cite{Agrawal2002}.

In the case of mixed input states, we find that if a Werner state is used as a resource, the difference between the classical fidelity and the maximal average fidelity is only a  monotonically increasing function in terms of the (increasing) entanglement of the resource state ($p\to1$) when the basis on which Alice measures is the Bell basis ($c_n=1$). In the other cases of correlated bases, the difference between the average fidelity (above) and the classical fidelity occurs for states with an intermediate degree of entanglement ($p\neq 1$) showing a relationship between the entanglement of the measurement basis and the entanglement of the resource state to obtain an optimal average fidelity.

Another aspect of this relationship becomes apparent when classical-quantum (Bell diagonal) states are used as a resource. Two different behaviors are observed depending on the type of state used as the resource of the protocol. We find a direct relationship between the symmetry of the correlations present in the classical-quantum state and the basis on which Alice performs her measurement. While the maximal average fidelity decreases as the correlations in the measurement basis do, it is strongly perturbed when the classical-quantum states have their correlations not aligned with the measurement basis. Finally, we find that the maximal fidelity of teleportation decreases with the measurement quantum correlations (i.e. $c_n$ tends to zero) and that the protocol corresponding to the computational basis ($c_n=0$) leads to a maximal average fidelity of QT lower than the classical one.

\section*{Acknowledgments}
\noindent The authors acknowledge Consejo Nacional de Investigaciones Cient\'ificas y T\'ecnicas (CONICET), Argentina, for financial support. D. G. B. has a fellowship from CONICET. DGB and MP are also grateful to Universidad Nacional de La Plata (UNLP), Argentina.

\section*{Appendix}

\noindent For completeness, here we provide the marginal Bloch vectors and the covariance matrices of the measurement operators $M_i$, see Eq.  \eqref{eq:measurementsvonNeu}, corresponding to the states $\ket{\phi^+_l}$, $\ket{\phi^-_l}$, $\ket{\psi^+_p}$, and $\ket{\psi^-_p}$, Eqs. \eqref{eq:agrawalstate1}-\eqref{eq:agrawalstate4}, have respectively the following marginal Bloch vectors and covariance matrices,
\begin{align*}
\vect{n}_1&=\begin{bmatrix}
0\\
0\\
\frac{1-r_l^2}{1+r_l^2}
\end{bmatrix}, & \vect{m}_1&=\begin{bmatrix}
0\\
0\\
\frac{1-r_l^2}{1+r_l^2}
\end{bmatrix}, & C_1&=\begin{bmatrix}
\frac{2r_l \cos \phi_l}{1+r_l^2} & \frac{2r_l \sin \phi_l}{1+r_l^2} & 0\\
\frac{2r_l \sin \phi_l}{1+r_l^2}& -\frac{2r_l \cos \phi_l}{1+r_l^2} & 0\\
0 & 0 & 1 \\
\end{bmatrix}, \\
\vect{n}_2&=\begin{bmatrix}
0\\
0\\
\frac{r_l^2-1}{1+r_l^2}
\end{bmatrix}, & \vect{m}_2&=\begin{bmatrix}
0\\
0\\
\frac{r_l^2-1}{1+r_l^2}
\end{bmatrix}, & C_2&=\begin{bmatrix}
\frac{-2r_l \cos \phi_l}{1+r_l^2} & \frac{-2r_l \sin \phi_l}{1+r_l^2} & 0\\
\frac{-2r_l \sin \phi_l}{1+r_l^2}& \frac{2r_l \cos \phi_l}{1+r_l^2} & 0\\
0 & 0 & 1 \\
\end{bmatrix},
\\
\vect{n}_3&=\begin{bmatrix}
0\\
0\\
\frac{1-r_p^2}{1+r_p^2}
\end{bmatrix}, & \vect{m}_3&=\begin{bmatrix}
0\\
0\\
\frac{r_p^2-1}{1+r_p^2}
\end{bmatrix}, & C_3&=\begin{bmatrix}
\frac{2r_p \cos \phi_p}{1+r_p^2} & \frac{-2r_p \sin \phi_p}{1+r_p^2} & 0\\
\frac{2r_p \sin \phi_p}{1+r_p^2} & \frac{2r_p \cos \phi_p}{1+r_p^2} & 0\\
0 & 0 & -1 \\
\end{bmatrix},
\\
\vect{n}_4&=\begin{bmatrix}
0\\
0\\
\frac{r_p^2-1}{1+r_p^2}
\end{bmatrix}, & \vect{m}_4&=\begin{bmatrix}
0\\
0\\
\frac{1-r_p^2}{1+r_p^2}
\end{bmatrix}, & C_4&=\begin{bmatrix}
\frac{-2r_p \cos \phi_p}{1+r_p^2} & \frac{2r_p \sin \phi_p}{1+r_p^2} & 0\\
\frac{-2r_p \sin \phi_p}{1+r_p^2} & \frac{-2r_p \cos \phi_p}{1+r_p^2} & 0\\
0 & 0 & -1 \\
\end{bmatrix}.
\end{align*}
Thus, we can take a simpler parametrization,  
\begin{align} \label{eq:agrawaldecomp}
\vect{n}_i&=\begin{bmatrix}
0\\
0\\
n_i
\end{bmatrix}, & \vect{m}_i&=\begin{bmatrix}
0\\
0\\
m_i
\end{bmatrix}, & C_i&=\begin{bmatrix}
\tilde{c}_i O^{11}_i & \tilde{c}_i O^{12}_i & 0\\
\tilde{c}_i O^{21}_i & \tilde{c}_i O^{22}_i & 0\\
0 & 0 & -\det O_i \\
\end{bmatrix}.
\end{align}
being $|n_i|=|m_i|$, $\tilde{c}_i^2+n_i^2=1$ with $\tilde{c}_i\geq 0$, and, finally, $O_i$ a rotation matrix 2$\times$2 if $k=1,2$ and an improper rotation if $k=3,4$. For example, for k=1, $n_1=(1-r_l^2)/(1+r_l^2)=m_1$, $\tilde{c}_1=2r_l/(1+r_l^2)$ and $$
O_1=\begin{bmatrix}
\cos \phi_l &  \sin \phi_l \\
\sin \phi_l &  -\cos \phi_l 
\end{bmatrix}	
$$ implying $\det O_1=-1$. When $l$ and $p$ are real numbers, $O_i$ are diagonal matrices. If $l=p=r_n$, $\tilde{c}_i=c_n=2r_n/(1+r_n^2)$ for all $i$.
\section*{References}
\bibliographystyle{unsrt}
\bibliography{library}{}

\end{document}